\documentclass[amsmath,amssymb,twocolumn,prl,showpacs,nofootinbib,floatfix,superscriptaddress]{revtex4-2}
\usepackage{soul}
\usepackage{amsmath}
\usepackage{graphicx}
\usepackage{dcolumn}
\usepackage{bm}
\usepackage{natbib}
\usepackage{tikz}
\usepackage{color}
\usepackage{subfigure}
\usepackage{longtable}
\usepackage{dcolumn}
\usepackage{etoolbox}
\usepackage{lmodern}
\usepackage{slantsc}
\usepackage{textcomp}
\usepackage{gensymb}
\usepackage{lipsum}
\usepackage{dblfloatfix}
\usepackage{multirow}
\usepackage[colorlinks, linkcolor={blue},citecolor={blue},breaklinks]{hyperref}
\usepackage{placeins} 

\begin{document}

\preprint{APS/123-QED}

\title{$\beta$-delayed proton pandemonium: A first detailed $^{31}$Cl($\beta p \gamma$)$^{30}$P decay scheme}

\author{T. Budner}
\email{budnerta@msu.edu}
\affiliation{Department of Physics and Astronomy, Michigan State University, East Lansing, MI 48824, USA}
\affiliation{Facility for Rare Isotope Beams, Michigan State University, East Lansing, MI 48824, USA}

\author{M. Friedman}
\affiliation{National Superconducting Cyclotron Laboratory, Michigan State University, East Lansing, MI 48824, USA}
\affiliation{Racah Institute of Physics, Hebrew University, Jerusalem, Israel 91904}

\author{L. J. Sun}
\affiliation{Facility for Rare Isotope Beams, Michigan State University, East Lansing, MI 48824, USA}
\affiliation{School of Physics and Astronomy, Shanghai Jiao Tong University, Shanghai 200240, China}

\author{C. Wrede}
\email{wrede@frib.msu.edu}
\affiliation{Department of Physics and Astronomy, Michigan State University, East Lansing, MI 48824, USA}
\affiliation{Facility for Rare Isotope Beams, Michigan State University, East Lansing, MI 48824, USA}

\author{B. A. Brown}
\affiliation{Department of Physics and Astronomy, Michigan State University, East Lansing, MI 48824, USA}
\affiliation{Facility for Rare Isotope Beams, Michigan State University, East Lansing, MI 48824, USA}

\author{D. P\'erez-Loureiro}
\affiliation{National Superconducting Cyclotron Laboratory, Michigan State University, East Lansing, MI 48824, USA}

\author{J. Surbrook}
\affiliation{Department of Physics and Astronomy, Michigan State University, East Lansing, MI 48824, USA}
\affiliation{Facility for Rare Isotope Beams, Michigan State University, East Lansing, MI 48824, USA}

\author{A. Adams}
\affiliation{Department of Physics and Astronomy, Michigan State University, East Lansing, MI 48824, USA}
\affiliation{Facility for Rare Isotope Beams, Michigan State University, East Lansing, MI 48824, USA}

\author{Y. Ayyad}
\affiliation{National Superconducting Cyclotron Laboratory, Michigan State University, East Lansing, MI 48824, USA}
\affiliation{IGFAE, Universidade de Santiago de Compostela, E-15782 Santiago de Compostela, Spain}

\author{D. W. Bardayan}
\affiliation{Department of Physics and Astronomy, University of Notre Dame, Notre Dame, IN 46556, USA}

\author{K. Chae}
\affiliation{Department of Physics, Sungkyunkwan University, Suwon 440-746, Republic of Korea}

\author{A. A. Chen}
\affiliation{Department of Physics and Astronomy, McMaster University, Hamilton, ON L8S 4L8, Canada}

\author{K. A. Chipps}
\affiliation{Physics Division, Oak Ridge National Laboratory, Oak Ridge, TN 37830-37831, USA}
\affiliation{Department of Physics and Astronomy, University of Tennessee, Knoxville, TN 37996, USA}

\author{M. Cortesi}
\affiliation{Facility for Rare Isotope Beams, Michigan State University, East Lansing, MI 48824, USA}

\author{B. Glassman}
\affiliation{Department of Physics and Astronomy, Michigan State University, East Lansing, MI 48824, USA}
\affiliation{National Superconducting Cyclotron Laboratory, Michigan State University, East Lansing, MI 48824, USA}

\author{M. R. Hall}
\affiliation{Department of Physics and Astronomy, University of Notre Dame, Notre Dame, IN 46556, USA}

\author{M. Janasik}
\affiliation{Department of Physics and Astronomy, Michigan State University, East Lansing, MI 48824, USA}
\affiliation{National Superconducting Cyclotron Laboratory, Michigan State University, East Lansing, MI 48824, USA}

\author{J. Liang}
\affiliation{Department of Physics and Astronomy, McMaster University, Hamilton, ON L8S 4L8, Canada}

\author{P. O'Malley}
\affiliation{Department of Physics and Astronomy, University of Notre Dame, Notre Dame, IN 46556, USA}

\author{E. Pollacco}
\affiliation{D\'epartement de Physique Nucl\'eaire, IRFU, CEA, Universit\'e Paris-Saclay, F-91191, Gif-sur-Yvette, France}

\author{A. Psaltis}
\affiliation{Department of Astronomy and Physics, Saint Mary's University, Halifax, NS B3H 3C3, Canada}

\author{J. Stomps}
\affiliation{Department of Physics and Astronomy, Michigan State University, East Lansing, MI 48824, USA}
\affiliation{National Superconducting Cyclotron Laboratory, Michigan State University, East Lansing, MI 48824, USA}

\author{T. Wheeler}
\affiliation{Department of Physics and Astronomy, Michigan State University, East Lansing, MI 48824, USA}
\affiliation{Facility for Rare Isotope Beams, Michigan State University, East Lansing, MI 48824, USA}

\date{\today}

\begin{abstract}
\begin{description}

\item[Background]
Positron decays of proton-rich nuclides exhibit large $Q$ values, producing complex cascades which frequently involve various radiations, including protons and $\gamma$ rays. Often, only one of the two is measured in a single experiment, limiting the accuracy and completeness of the decay scheme. An example is $^{31}$Cl, for which protons and $\gamma$ rays have been measured separately in detail but never with substantial sensitivity to proton-$\gamma$ coincidences.

\item[Purpose]
Provide detailed measurements of $^{31}$Cl $\beta$-delayed proton decay including $\beta$-$p$-$\gamma$ sequences, extract spectroscopic information on $^{31}$S excited states as well as their $\beta^+$ feedings, and compare to shell-model calculations.

\item[Methods] A fast fragmented beam of $^{31}$Cl provided by the National Superconducting Cyclotron Laboratory (NSCL) was deposited in the Gaseous Detector with Germanium Tagging (GADGET) system. GADGET’s gas-filled Proton Detector was used to detect $\beta$-delayed protons, and the Segmented Germanium Array (SeGA) was used to detect $\beta$-delayed $\gamma$ rays.

\item[Results] As many as 20 previously unobserved $\beta$-delayed proton transitions have been discovered in the present data, most of which populate excited states of $^{30}$P. The first detailed $^{31}$Cl($\beta p \gamma$)$^{30}$P decay scheme is presented, including updated $\beta$-delayed proton-decay energies and intensities, as well as several new $^{31}$S levels. Improved agreement is found with theoretical calculations of the Gamow-Teller strengths $B(\text{GT})$ for $^{31}$S excitation energies $7.5 < E_x < 9.5$ MeV.

\item[Conclusions]
The present work demonstrates that the ability to detect $\beta$-delayed protons and $\gamma$ rays in coincidence is essential to construct accurate positron decay schemes for comparison to theoretical nuclear structure calculations. In some respects, this phenomenon for $\beta$-delayed protons resembles the pandemonium effect originally introduced for $\beta$-delayed $\gamma$ rays.

\end{description}
\end{abstract}

\maketitle

\begingroup
\let\clearpage\relax
\section{I. Introduction}

Experimental measurements employing $\beta$-decay spectroscopy serve as sensitive probes for studying the structure of atomic nuclei located far from the valley of stability. Specifically, extracting the $B(\text{GT})$ values from experimental data allows for detailed comparison with shell-model theory. However, as $Q$ values increase, so do the number of possible transitions, fragmenting the $\beta$-decay intensity across many highly excited, closely spaced levels in the daughter nucleus. This reduces the experimental sensitivity to detect distinct photopeaks corresponding to primary $\gamma$-ray transitions from particular excited states, since the decay intensity is smeared across a broader range of energies. The problem is further exacerbated by the fact that $\gamma$-ray detection efficiency tends to decrease with increasing $\gamma$ energy. In the present work, we investigate an experimental case similar to this well-known ``pandemonium effect'' in the context of $\beta$-delayed proton emission \cite{Hardy1977}.  

The $\beta$-delayed $\gamma$ and proton decay of neutron-deficient $^{31}$Cl has been used to measure crucial resonances for constraining the thermonuclear rate of the $^{30}$P($p,\gamma$)$^{31}$S reaction at peak classical nova temperatures \cite{Kankainen2006,Saastamoinen2011,Saastamoinen2016,Bennett2016,Bennett2018,Budner2022}. Our present findings are based on data collected at NSCL using the GADGET system during the $^{31}$Cl $\beta$-delayed proton-decay measurement described in Ref. \cite{Budner2022}. This previous study focused on the astrophysical implications of one particular low-energy, $\ell=0$ resonance, while the present study provides an exhaustive discussion of the information-rich coincidence spectra from higher-energy protons and $\gamma$ decays from $^{30}$P excited states, which were obtained during NSCL Experiment 17024. Analysis of high-statistics proton-$\gamma$ coincidence data has revealed many previously unobserved transitions, as well as the identification of several new proton-unbound $^{31}$S levels. We present the most detailed study of $^{31}$Cl $\beta$-delayed proton decay to date, reveal the first-ever construction of a $^{31}$Cl($\beta p \gamma$)$^{30}$P decay scheme, and compare these experimental results to theoretical calculations. 

\nopagebreak
\section{II. Previous Spectroscopic Studies}

The 2.5-min half-life of $^{30}$P excludes its use as a target for direct-reaction experiments \cite{NuDat30P}, and the lack of high-intensity $^{30}$P beams at the astrophysically relevant low energies prohibits measuring the $^{30}$P($p,\gamma$)$^{31}$S reaction in inverse kinematics. However, various indirect methods have been used to study the proton-unbound states of $^{31}$S, including single-nucleon transfer \cite{Vernotte1999,Ma2007,Wrede2009,Irvine2013,Kankainen2017} and charge-exchange reaction measurements \cite{Wrede2009,Wrede2007,Parikh2011,Parikh2016}, as well as in-beam $\gamma$-ray \cite{Jenkins2005,Jenkins2006,Doherty2012,Doherty2014} and $\beta^+$ spectroscopy \cite{Kankainen2006,Saastamoinen2011,Saastamoinen2016,Bennett2016}. The context surrounding the status of $^{31}$S levels as of 2016 is summarized by Saastamoinen \textit{et al.} in Ref. \cite{Saastamoinen2016} and was extended upon by Bennett \textit{et al.}, whose original work in Ref. \cite{Bennett2018} details the findings of the largest $^{31}$Cl($\beta\gamma$)$^{31}$S dataset to date. Since then, further experimental results have been published on the reactions $^{32}$S($p,d$)$^{31}$S \cite{Setoodehnia2020,Burcher2022} and $^3$He($^{32}$S,$\alpha$)$^{31}$S  \cite{Sun2023}, as well as the radioactive decay $^{31}$Cl($\beta p$)$^{30}$P \cite{Budner2022}.

Proton emission from $^{31}$S levels populated by $^{31}$Cl $\beta^+$ decay was first observed by \"Ayst\"o \textit{et al.}, reporting up to eight different proton decay branches in Refs. \cite{Aysto1982,Aysto1983,Aysto1985}. Subsequent studies confirmed the observation of two $^{31}$Cl $\beta$-delayed proton decays, attributing the other transitions to contaminant decays \cite{Ognibene1996}. The first application of $^{31}$Cl $\beta^+$ decay as a probe of potentially crucial resonances for $^{30}$P($p,\gamma$)$^{31}$S was performed by Kankainen \textit{et al.} \cite{Kankainen2006}, identifying a total of about a dozen $\beta$-delayed proton decays in the process and qualitatively validating many of the same transitions observed by \"Ayst\"o \textit{et al.} Until now, the Saastamoinen doctoral dissertation boasted the highest-statistics $^{31}$Cl($\beta p \gamma$)$^{30}$P dataset ever collected \cite{Saastamoinen2011}, reporting 16 unique $^{31}$Cl $\beta$-delayed proton channels, whose energies and intensities have been adopted in the evaluated literature for $\beta$-delayed charged-particle decays \cite{Batchelder2020}.

Saastamoinen's thesis experiment used double-sided silicon strip detectors to measure proton decays and a single high-purity germanium (HPGe) crystal to detect $\beta$-delayed $\gamma$ rays. This experimental setup has the advantage of good proton-energy resolution and high detection efficiency for proton-decay energies $E_p>500$ keV. However, it suffers from large $\beta^+$ backgrounds, which obscure weak, low-energy, $\beta$-delayed proton decays. It also lacks the high angular coverage of an HPGe array, limiting the statistics necessary for detailed analysis of proton-$\gamma$ coincidence spectra. Ref. \cite{Saastamoinen2011} tentatively reports evidence that some of the $\beta$-delayed protons from $^{31}$Cl decay populate excited states of $^{30}$P, specifically citing the observation of proton-$\gamma$ coincidences between events in the 1.2-MeV proton group and 1.5-MeV $\gamma$ rays. However, without a high $\gamma$-ray detection efficiency, determining which excited states were populated in $^{30}$P was not possible. Thus, $\beta^+$ feedings to $^{31}$S excited states were reported under the assumption that all $\beta$-delayed proton decays populate the $J^{\pi}=1^+$ ground state of $^{30}$P. However, utilizing the high-intensity, radioactive beams produced at NSCL and the GADGET detection system \cite{Friedman2019}, which couples a gaseous Proton Detector with the large solid-angle coverage of SeGA \cite{Mueller2001}, we have collected the largest $^{31}$Cl $\beta$-delayed, proton-$\gamma$ coincidence dataset ever measured. 
\section{III. NSCL Experiment 17024}

A stable beam of $^{36}$Ar was extracted from an electron cyclotron resonance ion source and accelerated to an energy of 150 MeV per nucleon using the NSCL Coupled Cyclotron Facility. At an ion current of 75 pnA, this primary beam was impinged on a 1645-mg/cm$^2$-thick beryllium fragmentation target, producing a cocktail beam of various nuclides. Using magnetic rigidity and a 150-mg/cm$^2$-thick aluminum wedge, the A1900 fragment separator was optimized for maximal $^{31}$Cl beam intensity while minimizing impurities from other nuclear species \cite{Stolz2005}. The secondary beam was further purified using the Radio Frequency Fragment Separator (RFFS), which utilizes a time-varying electric field synchronized with the cyclotron RF \cite{Bazin2009}. The phase of this oscillating field pulse was tuned to selectively deflect the ions of interest through the narrow exit slits, while simultaneously filtering out undesirable contaminants. Upon exiting the RFFS, a 65\% pure beam of $^{31}$Cl was delivered to the experimental vault at an average rate of $6000$ pps. In decreasing order of intensity, the main beam contaminants identified in the $\Delta E$-TOF spectrum were the isotones $^{28}$Si, $^{30}$S, and $^{29}$P, as shown in Fig. \ref{fig:particleID}. Fortunately, none of these species are $\beta$-delayed charged-particle emitters that would contribute to background in our proton spectrum.

\begin{figure}
\centering
\includegraphics[scale=0.48]{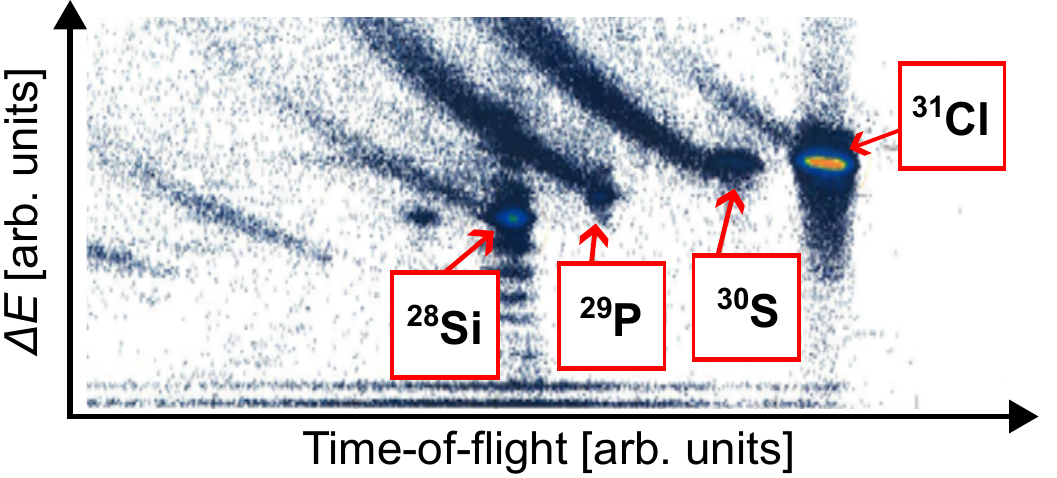}
\caption{\label{fig:particleID} Particle identification spectrum for the $^{31}$Cl cocktail beam. The energy losses $\Delta E$ of beam ions as measured by a silicon PIN detector are plotted against their time-of-flight (TOF) as measured between two plastic scintillators in the A1900 Fragment Separator.}
\end{figure}

A six-sided, cross-shaped vacuum chamber was connected to the end of the radioactive beam line. This diagnostic cross housed a retractable 300-$\mu$m-thick PIN detector, whose position within the chamber was controlled remotely using a pneumatic drive. Between data production runs, the beam current was attenuated and the silicon PIN diode was inserted into the path of the beam in order to measure the energy loss $\Delta E$ of ions passing through the semiconducting material. After collecting PIN data for $\approx10$ min, the relative peak positions corresponding to the different ions within the $\Delta E$ histogram were compared with \textsc{LISE++} calculations of $\Delta E$ in 300 $\mu$m of silicon for all nuclear species present in the secondary beam \cite{Tarasov2008}. Along with the expected $\gamma$-ray signatures observed in SeGA, this particle identification method consistently verified that $^{31}$Cl and $^{28}$Si comprised the majority of nuclear fragments delivered during the course of the $\beta$-decay measurement.

Before resuming each data production cycle, the PIN detector was retracted from the beam line so that the unattenuated ions could be implanted in the center of the GADGET setup, thermalizing in a cylindrical volume of 800-Torr, P10 gas within the Proton Detector chamber. To stop the bulk of the radioactive beam in the middle of the gaseous detector's active region, a rotatable, 0.75-mm-thick degrader foil of ultra-pure ($>99.99\%$) aluminum was used to tune the energy of the incoming $^{31}$Cl particles. This beam-energy degrader was positioned in the air gap between the end of the beam line and the entrance aperture to the detector chamber, which was sealed with a thin window of Kapton polyimide film. The degrader angle was measured using a circular protractor and was manually adjusted until the $^{31}$Cl distribution was longitudinally centered within GADGET, as shown in Fig. \ref{fig:longBeamDist}.

The Proton Detector's principle of operation was based on the AstroBox system developed at Texas A\&M University \cite{Pollacco2013}. GADGET's conceptual design, performance testing, and successful commissioning experiment at NSCL are thoroughly described in Ref. \cite{Friedman2019}. In short, the calorimetric design of the Proton Detector was built to measure weak, low-energy, $\beta$-delayed charged-particle decays with high efficiency and to suppress $\beta$ backgrounds. The slower, more massive protons ionize the dense fill gas and stop over short ranges within the detector's cylindrical active region, while the lighter, faster-moving $\beta$ particles escape the active region, ionizing very little of the gas in the process. A uniform, 150-V/cm electric field ensures that the drift time for the ionization electrons produced by $\beta$-delayed charged-particle radiation is proportional to the distance between the location of the $\beta$-decay event and the Proton Detector's anode plane. Under these conditions, the maximum drift time for electrons traversing a distance between the cathode, which supplies the voltage for the drift field, and the detector's amplification region within the Micro-Mesh Gaseous Structure (MICROMEGAS) is $\approx7.5$ $\mu$s \cite{Giomataris1996}. This defines a 7.5-$\mu$s time window after a $\gamma$-ray event is detected by SeGA for a coincident charged-particle event to be observed in the Proton Detector. 

\begin{figure}
\centering
\includegraphics[width=\linewidth]{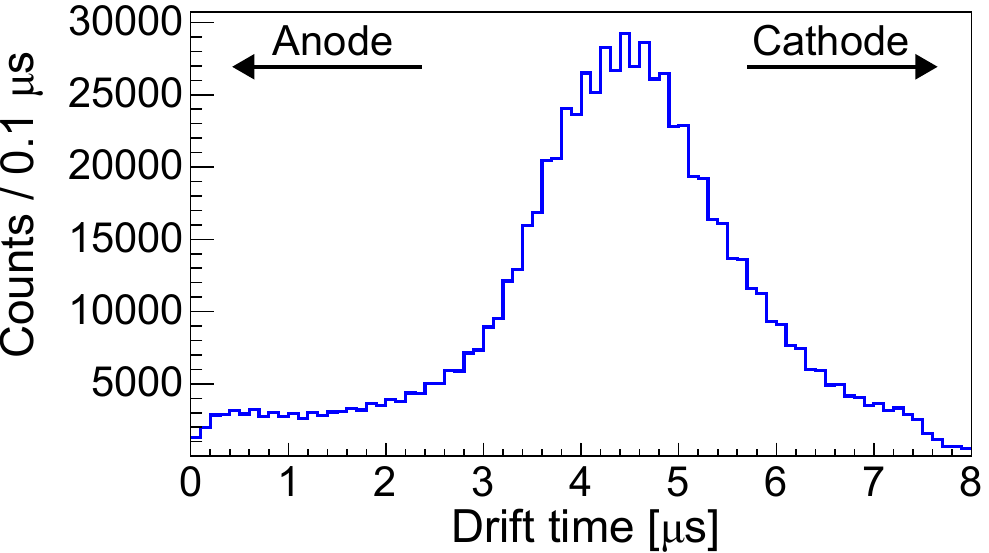}
\caption{\label{fig:longBeamDist} Drift times of ionization electrons produced by $\beta$-particle events, for which $E_{\beta}<200$ keV, detected in coincidence with 2234-keV $\gamma$ rays. This serves as a rough approximation of the longitudinal distribution for $^{31}$Cl $\beta^+$ decays throughout the active volume of the Proton Detector.}
\end{figure}

\begin{figure}
\centering
\includegraphics[scale=0.52]{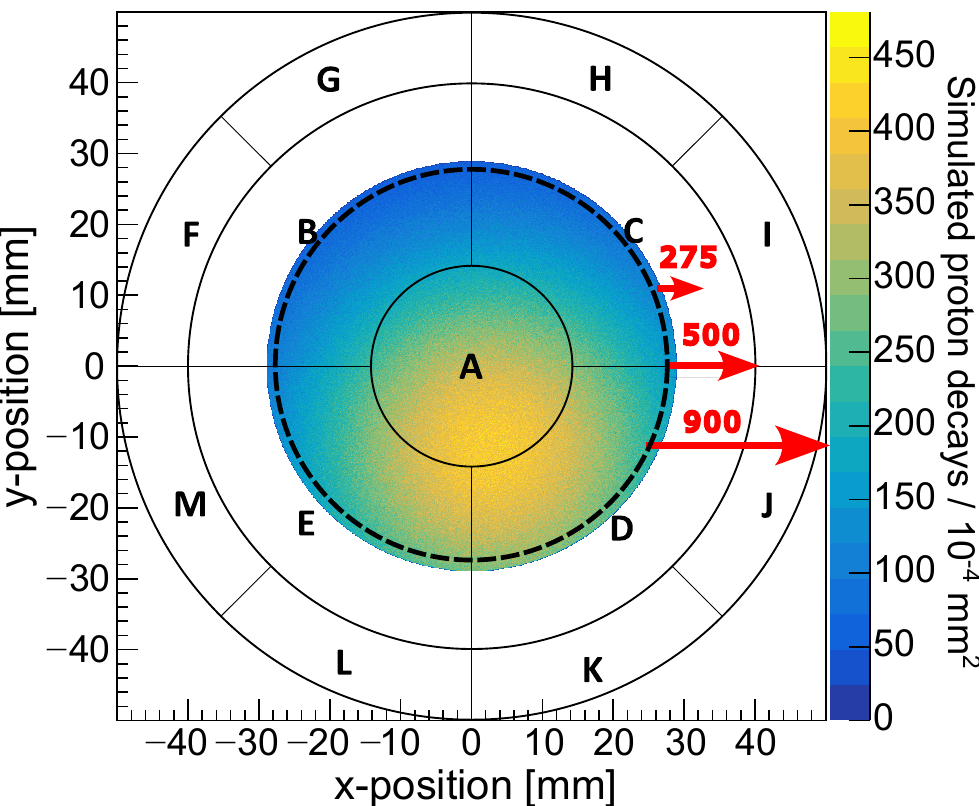}
\caption{\label{fig:beamSpot} Reconstructed beam spot representing the transverse distribution of $^{31}$Cl ions overlaid on the segmented anode plane. Active detector pads are labeled A--E, and the outer veto pads are labeled  F--M. The dashed circle shows the projection of the beam entrance aperture onto the MICROMEGAS board. Arrow sizes correspond to the track lengths of ionization electrons produced by 275-, 500-, and 900-keV proton decays constrained to the $x$-$y$ plane in an 800-Torr, P10 fill gas, according to the Stopping and Range of Ions in Matter (\textsc{SRIM}) calculator \cite{Ziegler2010}.}
\end{figure}

As depicted in Fig. \ref{fig:beamSpot}, the anode detector plane is segmented into 13 charge-sensitive MICROMEGAS pads, which are responsible for providing gain to the electronic signals produced by $\beta$-decay events. The five inner pads are used to measure the full energy and intensity of $\beta$-delayed proton decays, while the eight outer pads are used to veto higher-energy protons that would otherwise register cumulatively as a continuum of partial energy depositions in the spectrum. Decay events whose ionization tracks deposit energy in any of the veto pads are rejected from our analysis to reduce continuum background in the $\beta$-delayed proton spectrum. The active region of the Proton Detector is contained within a cylindrical volume between the anode and cathode planes, whose radius is defined by the boundary between the inner active pads and the outer veto pads.

For rare isotope beam experiments, GADGET operates in alternating cycles. During the beam-implantation phase, radioactive nuclei accumulate in the gaseous detector chamber. This is followed by a decay-measurement period, during which beam delivery is turned off, and $\beta$-delayed proton decays are detected by the MICROMEGAS. Given the 190-ms half-life of $^{31}$Cl \cite{Saastamoinen2011}, a 300-ms implantation period and a 200-ms measurement window were chosen to maximize detection efficiency for $\beta$-delayed proton decay, while still protecting the MICROMEGAS preamplifier channels during beam implantation. The Proton Detector employs an electronic gating grid of gold-plated copper wires in front of the anode plane. Its applied bias voltage is synchronized with the beam-delivery cycle, such that strong ionization currents induced by ion implantation are repelled, while weak, low-energy, $\beta$-delayed proton decays are still detectable during the 200-ms measurement window.

The energies and timestamps of events detected by the diagnostic PIN detector, the 13 charge-sensitive MICROMEGAS pads, the central contacts of all 16 SeGA crystals, as well as the switch signal responsible for synchronizing the gating grid with the beam cycle, were collected using two 16-channel PXI digital processor cards and the NSCL Digital Data Acquisition System \cite{Prokop2014}. The cumulative dataset referenced throughout this article consists of measurements taken over the course of $\approx86$ h of radioactive beam time, with most data production runs lasting $\approx 60$ min each.

\section{IV. Data Analysis}

\subsection{A. $\gamma$-ray energy and efficiency calibrations}

The $\gamma$-ray singles spectra were gain-matched on a run-by-run basis using four well-known room-background $\gamma$ rays from $^{228}$Ac, $^{40}$K, $^{214}$Bi, and $^{208}$Tl, with energies of 911, 1461, 1764, and 2615 keV, respectively \cite{Gilmore2008}. A single $^{31}$Cl $\beta$-delayed $\gamma$ transition at 6279 keV was also used to extend this calibration to higher energies. In the uncalibrated spectra, these photopeaks were fit using Gaussian distributions for each SeGA detector. Linear relationships between the Gaussian centroid positions and their corresponding $\gamma$ energies were determined for 15 of the 16 detectors; a single HPGe crystal exhibited poor energy resolution and was excluded from this analysis. 

Enforcing the condition that $\gamma$ rays must be observed in coincidence with either $\beta^+$ or proton events detected by the MICROMEGAS results in the spectrum shown in Fig. \ref{fig:betaGatedGammas}. Gating on Proton Detector events when plotting the $\gamma$ spectrum not only reduces the amount of room background observed in SeGA, but it also removes $\gamma$ events originating from $^{31}$Cl decays which occurred outside the ionization chamber or during the beam implantation period when the electronic gating grid prevents signal collection in the MICROMEGAS. This is useful for the purpose of normalization when determining relative decay intensities.

\begin{figure}
\includegraphics[width=\linewidth]{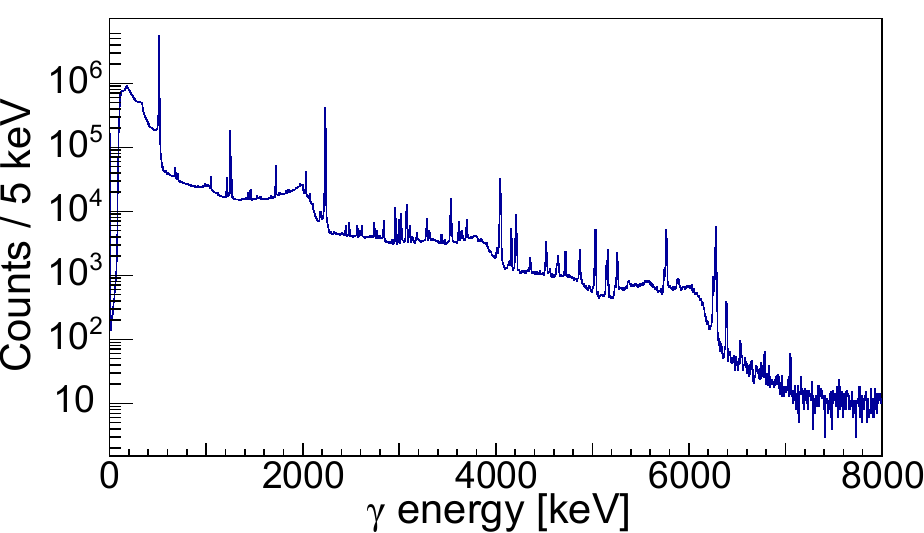}
\caption{The $^{31}$Cl $\beta$-delayed, $\gamma$-ray singles spectrum, gated on all coincident charged-particle detections, including $\beta^+$ particles. Any $\gamma$ events recorded in this spectrum must be observed within the same 7.5-$\mu$s window as a charged-particle event detected in the Proton Detector.}
\label{fig:betaGatedGammas}
\end{figure}

GADGET's efficiency to detect $\gamma$ rays was determined for energies $E_{\gamma}\lesssim7$ MeV. This was achieved by first measuring the number of observed $\gamma$-ray transitions across a range of energies. Out of the strongest $\gamma$-ray transitions from $^{31}$S excited states, 12 well-separated photopeaks were fit with asymmetric Gaussian distributions, while the local backgrounds were modeled as either first- or second-degree polynomials.  The exponentially modified Gaussian (EMG) distribution used to model each spectral peak can be written in terms of the complementary error function (erfc) as

\begin{equation}
\begin{split}
   f(x;\mu,\sigma,\lambda) = & \frac{N\lambda}{2}\text{exp}
\left[ \frac{\lambda}{2} \left(\lambda\sigma^2 \pm2\mu \mp2x \right) \right] \\
& \times
\text{erfc} \left( \frac{\lambda\sigma^2 \pm\mu \mp x}{\sqrt{2}\sigma}\right), 
\end{split}
\label{eq:EMG}
\end{equation}

\noindent
where $\mu$ and $\sigma$ correspond to the centroid and width of the function's Gaussian component, respectively, while $\lambda$ determines the size of the asymmetry. The plus-minus signs imply that this response function can account for skew in both directions. An asymmetric distribution with a high-energy tail was used for the Proton Detector response function to account for $\beta^+$ summing, while an EMG with a low-energy tail was used to model incomplete charge collection in the SeGA detectors. The normalization factor $N$ is the definite integral of the function, which determines the total number of recorded decay events in each peak.

Since Ref. \cite{Bennett2018} is the most complete study of $^{31}$Cl($\beta\gamma$)$^{31}$S to date, we adopt its $\beta$-decay feedings $I_{\beta}$, which were used to deduce the corresponding $\gamma$ intensities $I_{\gamma}$. However, the initial $I_{\gamma}$ values reported by Bennett \textit{et al.} contained some minor inconsistencies and have since been corrected in the evaluated literature \cite{NuDat31S}. These corrected values were used to determine SeGA's relative detection efficiency as a function of $\gamma$-ray energy, as shown in Fig. \ref{fig:SeGAefficiency}. 

\begin{figure}
    \centering
    \includegraphics[width=\linewidth]{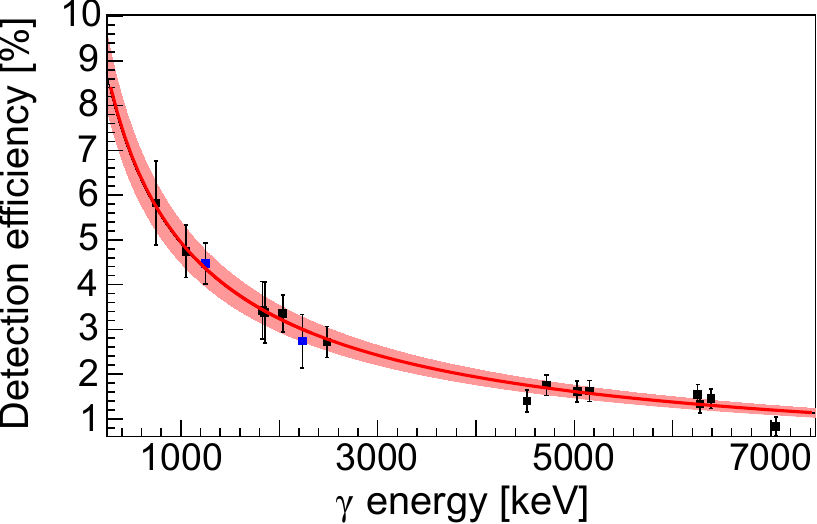}
    \caption{Absolute detection efficiency for SeGA plotted with respect to $E_{\gamma}$. Relative efficiencies were evaluated at energies marked by black data points and were fit with a power-law function (red online). The relative efficiency curve was then scaled to intersect the two fixed markers (blue online), whose absolute efficiencies were evaluated at 1248 and 2234 keV.}
    \label{fig:SeGAefficiency}
\end{figure}

To find the probability of detecting a $\gamma$-ray event with energy $0.6 < E_{\gamma} < 7.1$ MeV, the fit in Fig. \ref{fig:SeGAefficiency} was scaled to match the absolute detection efficiencies of the two strongest $\gamma$ transitions in the spectrum, which were independently determined. Normalizing detection efficiency at specific energies is possible if two quantities are well known: (1) the observed number of incoming transitions $\gamma_{\text{in}}$ for a given level, as determined by fitting the $\gamma$-ray singles spectrum in Fig. \ref{fig:betaGatedGammas}, and (2) the number of outgoing transitions $\gamma_{\text{out}}$ detected in coincidence with $\gamma_{\text{in}}$, as in Fig. \ref{fig:hGammaGammaCoincidencesSeGAefficiency}, for example. Assuming there are no other outgoing $\gamma$ transitions from this level, we can express the probability of detecting $\gamma_{\text{out}}$ as the ratio 

\begin{equation}
\epsilon(\gamma_{\text{out}}) = \frac{\text{Observed $\gamma_{\text{in}}$-$\gamma_{\text{out}}$ coincidences}}
{\text{Observed $\gamma_{\text{in}}$} \text{ singles}}.
\label{eq:absEff}
\end{equation}

\noindent
For the present setup, this ratio was scaled by a factor of 15/14 to account for possible coincidence photons summing their combined energy in a single SeGA detector. This relationship was used to determine the absolute detection efficiencies of 4.5(5)\% and 2.7(6)\% for the 1248- and 2234-keV $\gamma$ rays, respectively. These correspond to ground-state transitions from the two lowest-lying excited states in $^{31}$S. The fact that these levels are populated via several intense $\gamma$ decays from more highly excited states allowed us to verify our method of absolute efficiency evaluation using multiple incoming $\gamma$ transitions. 

\begin{figure}
    \centering
    \includegraphics[width=\linewidth]{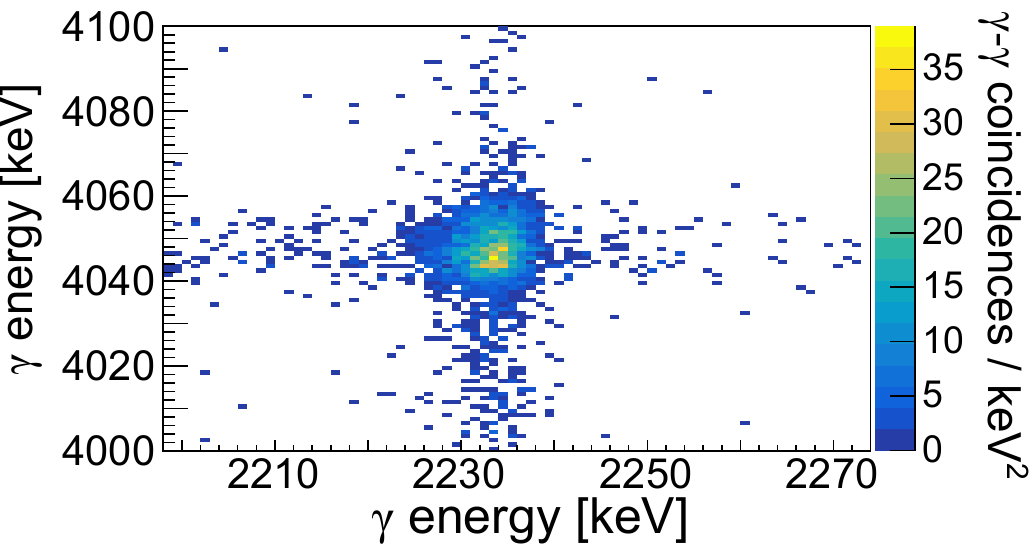}
    \caption{Observed $\gamma$-$\gamma$ coincidences between the 4045- and 2234-keV transitions in the de-excitation cascade from the strongly populated 6279-keV isobaric analog state. Employing Eq. \ref{eq:absEff}, this pair of transitions was used to normalize the absolute $\gamma$-ray detection efficiency at 2234 keV for SeGA.}
    \label{fig:hGammaGammaCoincidencesSeGAefficiency}
\end{figure}

We do not find evidence for additional $^{31}$S excited state decays via $\gamma$-ray emission beyond what has been reported in Ref. \cite{Bennett2018}. However, a detailed analysis of proton-$\gamma$ coincidence data reveals many previously unobserved $\beta$-delayed proton decays, most of which populate excited states of $^{30}$P. Knowing the absolute SeGA efficiency over a wide range of $\gamma$ energies is useful for determining the intensity of weak, $\beta$-delayed proton decays detected in coincidence with $\gamma$ rays emitted from $^{30}$P excited states.

\subsection{B. Proton-$\gamma$ coincidences and proton energy calibration}

\begin{figure}[t]
\includegraphics[width=\linewidth]{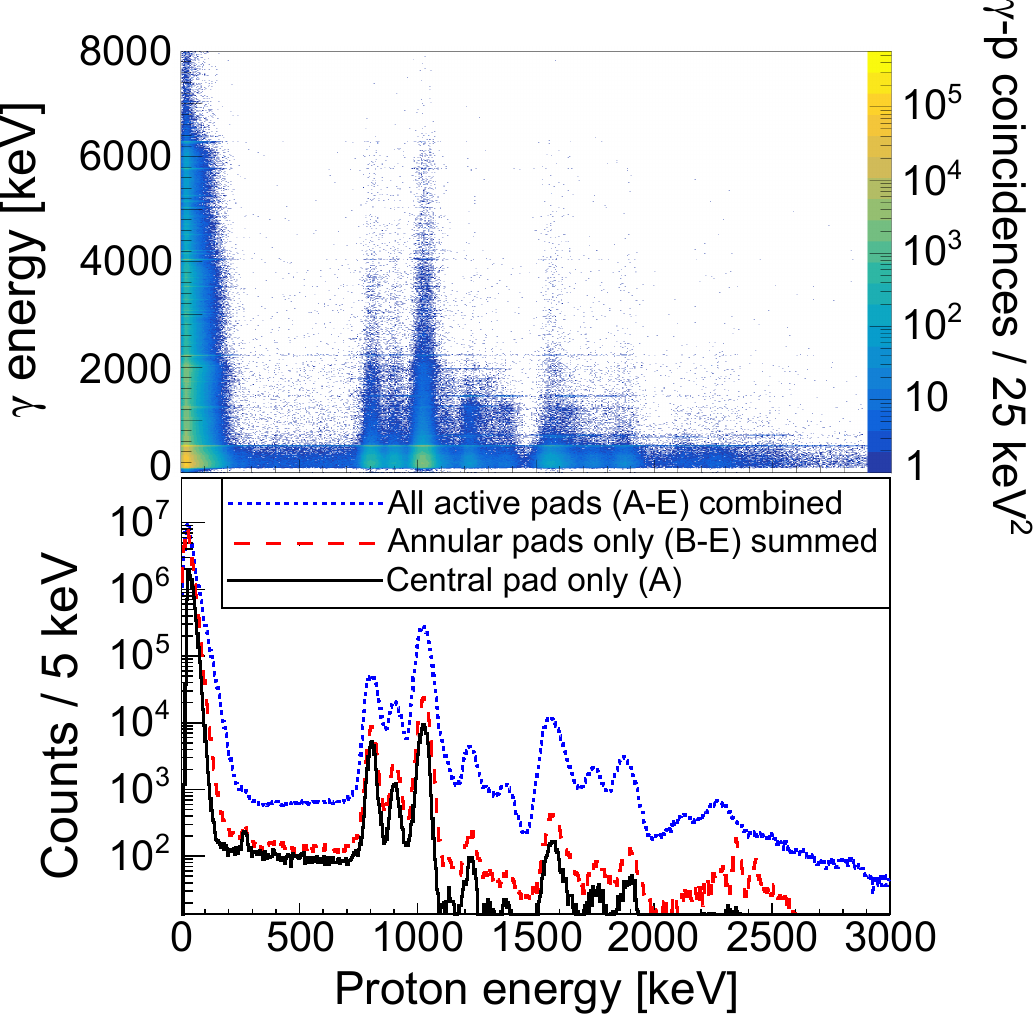}
\caption{(Top) All observed coincidences between $\gamma$ rays and charged-particle decays, including $\beta$-delayed protons. (Bottom) Proton Detector singles spectra for all non-vetoed events measured in the active region [dotted gray (indigo online)], the summed statistics from each of the four annular pads' individual spectra [dashed gray (red online)], and decay events only detected in the central MICROMEGAS pad [solid black].}
\label{fig:protonSpectra}
\end{figure}

As with the $\gamma$-ray data, the proton spectra were gain-matched on a run-by-run basis for each of the five inner MICROMEGAS pads. However, because there were significantly fewer proton counts per run than $\gamma$-ray events, Proton Detector data acquired over shorter runs ($\lesssim 60$ min) were combined with events from longer, consecutive runs. The limited statistics also meant that this linear gain-matching procedure relied on the positions of the only two peaks that could be consistently fit with Gaussian distributions in the spectra of all active pads. These features correspond to the two strongest $^{31}$Cl $\beta$-delayed proton transitions, with decay energies of 0.8 and 1.0 MeV \cite{Saastamoinen2011,Saastamoinen2016,Batchelder2020}.

The singles spectra are shown in the bottom panel of Fig. \ref{fig:protonSpectra} for all non-vetoed, $\beta$-delayed proton decay events observed in the active region of the Proton Detector, events observed only in the central pad, as well as the summed statistics of annular pads B--E where events were detected in only one of the four pads. The individual pad spectra have good energy resolution, but the lack of statistics causes gain matching at higher energies to be unreliable, resulting in the separation of the 2.3-MeV peak position across the various detector pads. Conversely, while the cumulative event-level energy-summing spectrum exhibits comparatively worse energy resolution, it has the advantage of higher statistics, more robust gain matching, and a well-defined efficiency model. Therefore, the latter is more useful for extracting a precise energy calibration and determining $\beta$-delayed proton decay intensities.

For energies $E_p<300$ keV, our Proton Detector spectra are dominated by exponential backgrounds from $\beta^+$ particles. However, we are able to resolve a small proton peak corresponding to the decay of a crucial $\ell=0$ resonance for the $^{30}$P($p,\gamma$)$^{31}$S reaction in oxygen-neon novae, which is the subject of Ref. \cite{Budner2022}. In the energy region $300<E_p<700$ keV, the relatively flat background results from a ``wall effect,'' whereby higher-energy protons near the upstream or downstream end of the detector chamber deposit some, but not all, of their energy into ionizing the fill gas before implanting in the cathode or anode, respectively.

\begin{figure}[t]
\centering
\includegraphics[width=\linewidth]{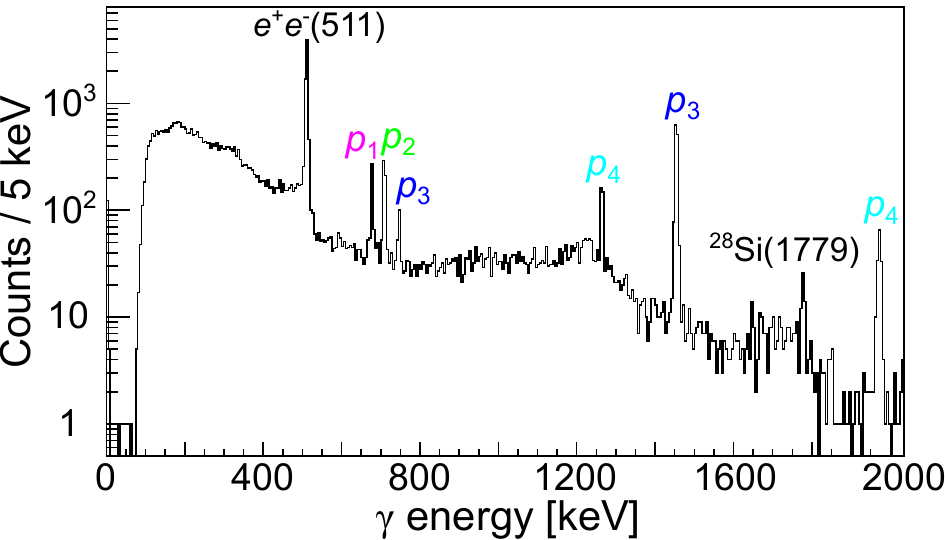}
\caption{\label{fig:protonGatedGammas} Energy spectrum for $\gamma$ rays detected in coincidence with proton decays in the energy region $1.2< E_p\text{(c.m.)}< 1.3$ MeV. The photopeaks labeled $p_1$, $p_2$, $p_3$, and $p_4$ correspond to $\gamma$ transitions from the first, second, third, and fourth excited states of $^{30}$P, respectively, which are populated by $^{31}$Cl $\beta$-delayed proton decays. Also labeled is the 511-keV annihilation photopeak and $\gamma$ rays from the decay of the first excited state of $^{28}$Si, which is populated by $\beta$-delayed proton decay of the beam contaminant $^{29}$S \cite{Viera1979,Zha2026}.}
\end{figure}

For center-of-mass energies $E_p\text{(c.m.)}>800$ keV, we observe almost all of the same $^{31}$Cl $\beta$-delayed proton decay peaks previously reported in literature \cite{Aysto1982,Aysto1983,Aysto1985,Ognibene1996,Kankainen2006,Saastamoinen2011,Saastamoinen2016}. However, due to the lack of high-statistics, proton-$\gamma$ coincidence information, these $\beta$-delayed proton decays were assumed to be direct  transitions to the ground state of $^{30}$P \cite{Batchelder2020}. The $\gamma$-tagging capabilities of GADGET reveal that certain features resembling individual peaks within the proton singles spectra actually contain multiple contributions from various $\beta$-delayed proton transitions feeding different $^{30}$P levels. This is illustrated in Fig. \ref{fig:protonGatedGammas}, which plots only the $\gamma$-ray events detected in coincidence with proton decays for which 1.2 $< E_p\text{(c.m.)}<$ 1.3 MeV. In this proton-gated $\gamma$ spectrum, there are six photopeaks that clearly stand out from the background, whose energies correspond to all known $\gamma$ decays of the first, second, third, and fourth excited states of $^{30}$P \cite{Ray2007,Ramstein1981,Vermeulen1979,Moss1971}. This implies direct population from $^{31}$Cl $\beta$-delayed proton transitions $p_1$, $p_2$, $p_3$, and $p_4$, respectively.

Conversely, by placing coincidence gates on $\gamma$ rays emitted from all energetically accessible excited states of $^{30}$P, we are able to plot a unique spectrum for each final state populated by $\beta$-delayed proton decay in the daughter nucleus. For example, Fig. \ref{fig:hLevel5protons} clearly shows an excess of protons near 740 keV as well as another potential $p_5$ peak above background at $E_p\approx1060$ keV, suggesting that the fifth excited state at 2539 keV is also populated by $^{31}$Cl $\beta$-delayed proton decay. 

\begin{figure}[t]
\centering
\includegraphics[width=\linewidth]{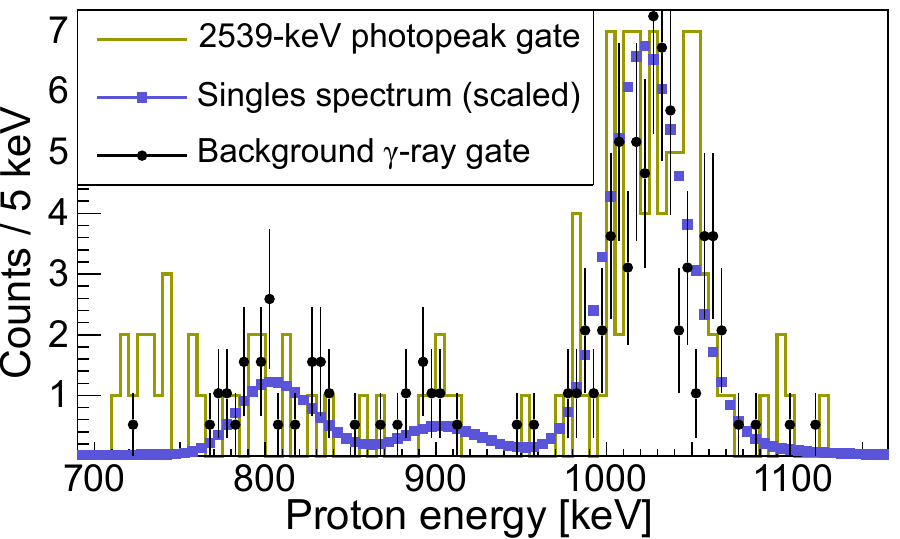}
\caption{\label{fig:hLevel5protons} The energy spectrum of protons in coincidence with 2539-keV $\gamma$ rays is plotted as a solid-line histogram (gold online) in comparison to two different background models. The square markers (indigo online) show a scaled down singles spectrum, representing a random distribution of proton coincidences, while the circular black markers show protons in coincidence with $\gamma$ events sampled from the background region $2660 < E_{\gamma} <2700$ keV.}
\end{figure}

Careful background modeling was performed in order to account for both the ``accidental'' contributions of proton transitions in random coincidence with $^{30}$P $\gamma$ rays, as well as the ``real'' contributions from non-random coincidences between protons populating  $^{30}$P excited states and the $\gamma$ rays that fall within the low-energy tail of the corresponding transition's photopeak. Random background coincidences were modeled using the all-pads singles spectrum shown in Fig. \ref{fig:protonSpectra} scaled down to the statistics of the $\gamma$-gated proton spectrum. Since the 1.0-MeV proton decay is the strongest in the spectrum and corresponds to a ground-state ($p_0$) transition, the scaling factor was defined by the ratio between the total number of 1.0-MeV singles and the number of 1.0-MeV protons observed in the $\gamma$-gated spectrum. The background contribution from potentially non-random coincidences was approximated using the $\gamma$-gated proton spectrum sampled from a relatively featureless part of the spectrum, slightly higher in energy than the photopeak-sampled region with comparable statistics. Figure \ref{fig:hLevel5protons} illustrates both of these background models in the case of $p_5$-$\gamma$ coincidences.

\begin{figure}
\centering
\includegraphics[width=\linewidth]{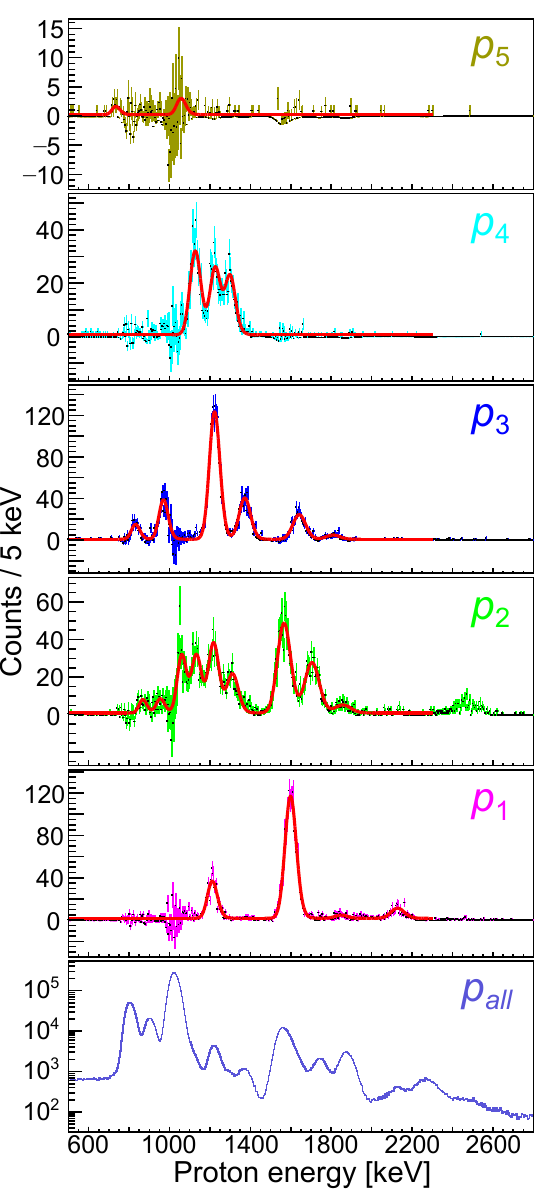}
\caption{\label{fig:backgroundSubtractionFits} Comparison between the $p_{\text{all}}$ singles spectrum (indigo online) and the $\gamma$-gated proton spectra for each of the first five excited states of $^{30}$P after background subtraction. The $p_1$ (magenta online), $p_2$ (green online), $p_3$ (blue online), $p_4$ (cyan online), and $p_5$ (gold online) spectra show the $\beta$-delayed protons populating the 677-, 709-, 1454-, 1973-, and 2539-keV levels, respectively. The superimposed fits (red online) were used to determine the decay energies and intensities of these proton transitions.}
\end{figure}

This background subtraction procedure results in the $\gamma$-gated proton spectra depicted in Fig. \ref{fig:backgroundSubtractionFits}, which clearly shows new transitions populating the 677-, 709-, 1454-, 1973-, and 2539-keV excited states of $^{30}$P, whose spin-parity assignments for these levels are $J^{\pi} = 0^+$, $1^+$, $2^+$, $3^+$, and $3^+$, respectively \cite{Boerma1975,NuDat30P}. Many of these previously unidentified $\beta$-delayed proton decays are quite weak and would otherwise be obscured in the ungated singles spectrum. In energy regions where there are strong $p_0$ transitions, such as around 1.0 MeV, the uncertainties of the bin-by-bin subtraction between the photopeak-gated and background coincidence histograms are relatively large, making it difficult to identify particularly weak $\beta$-delayed proton decays in this energy regime. In addition to these statistical uncertainties, the difference between the two background subtraction methods along with variations in modeling peak shape as a function of $E_p$ result in systematic uncertainties associated with fitting the  spectra in Fig. \ref{fig:backgroundSubtractionFits}. For well-defined and reasonably sized $\gamma$-gated proton peaks, the relative effects on these systematics are within the range of $1-10 \%$, but for very weak protons, the relative uncertainties are significantly larger ($\lesssim30\%$). Therefore, for the latter cases, we can only assign upper limits on their proton-decay intensities.

The identification of many new $\beta$-delayed proton decays in Fig. \ref{fig:backgroundSubtractionFits} allows us to refine the energy calibration from Ref. \cite{Budner2022} and extend it to higher proton-decay energies. To do this, we first adopted the evaluated level energies from the latest version of the $A=31$ Nuclear Data Sheets for the three strongest proton-emitting states populated in $^{31}$Cl $\beta^+$ decay at 6936(2), 7037(2), and 7157.5(11) keV \cite{NuDat31S}. Utilizing the relationship

\begin{equation}
E_{x_i}(^{31}\text{S}) = S_p(^{31}\text{S}) + E_p(\text{c.m.}) + E_{x_f}(^{30}\text{P}),
\label{eq:exEn}
\end{equation}

\noindent
where $E_{x_i}$ is the initial excitation energy, $S_p=6130.65(24)$ keV is the $^{31}$S proton separation energy \cite{Wang2021}, and $E_{x_f}$ is the excitation energy of the final $^{30}$P state, we computed the center-of-mass decay energies of our initial calibration points $E_p(\text{c.m.})$: 805(2), 906(2) and 1026.9(11) keV. By fitting the corresponding proton peaks in the combined singles spectrum of Fig. \ref{fig:protonSpectra} and determining a linear relationship between peak position and decay energy, we defined a local calibration with a 3-keV uncertainty band in the region $0.8 < E_p < 1.0 $ MeV.

In Fig. \ref{fig:backgroundSubtractionFits}, we identified several weak, $\beta$-delayed protons within our valid calibration region, including one particular decay at $E_{p_3}=834(3)$ keV, which populates the 1454-keV level of $^{30}$P. Again using Eq. \ref{eq:exEn}, we determined the $^{31}$S excitation energy of the corresponding proton-unbound state to be $E_{x_i}=8418(3)$ keV. This is quite close to a previously observed level, whose evaluated excitation energy $E_{x_i}=8424(3)$ keV \cite{NuDat31S} is a weighted average of the results from Refs. \cite{Wrede2009,Saastamoinen2011,Setoodehnia2020}. Thus, we consider the present work to be an independent measurement of this level and use our 8418-keV excitation energy for the purpose of our calibration.

Further analysis revealed that this 8418-keV level is responsible for three additional proton peaks observed in our spectra, one of which is likely the same 2.3-MeV proton transition reported in Ref. \cite{Saastamoinen2011}. The other two $\beta$-delayed proton transitions through this state occur at decay energies near 1.6 MeV and populate the 677- and 709-keV excited states of $^{30}$P. Thus, four well-known $^{31}$S levels can be used to constrain seven proton-energy calibration points across the region $0.8 < E_p < 2.3$ MeV, making linear extrapolation to lower decay energies more robust. Applying this more precise calibration to evaluate the energy of the astrophysically relevant resonance in Ref. \cite{Budner2022}, we report a revised value of $E_{p_0}=258(4)$ keV with reduced uncertainty, bringing us into agreement with Bennett \textit{et al.} at the level of a single standard deviation \cite{Bennett2016,Bennett2018}.

The energies of all $^{31}$Cl $\beta$-delayed proton decays observed in NSCL Experiment 17024 are provided in Table \ref{table:allProtonData}. Equation \ref{eq:exEn} allowed us to use these calibrated proton-decay energies to compute the initial excitation energies, which are also tabulated. In other cases where multiple proton transitions are attributed to the decay of the same $^{31}$S level, a weighted average is used for the excitation energy.

\subsection{C. Proton intensities and efficiency simulations}

The high-statistics, proton-$\gamma$ coincidence information obtained with GADGET allows us to place stricter constraints when analyzing the cumulative proton energy spectrum across all active detector pads shown in Fig. \ref{fig:protonSpectra}. The proton singles were fit using multiple EMG distributions to model proton peaks across four energy regimes: $0.7<E_p<1.3$ MeV, $1.1<E_p<1.4$ MeV, $1.5<E_p<1.8$ MeV, and $2.0<E_p<2.7$ MeV. The local backgrounds were modeled as linear, and the EMG response function describing the peak shape was constrained to vary smoothly with decay energy, under the assumption that all resonances are intrinsically narrow; no evidence of broad resonances was identified.

Initially, the coarse features of the cumulative proton singles spectrum were fit under the assumption that each visible peak corresponds to a unique $p_0$ decay, allowing the Gaussian centroid and integral of the distribution to vary freely in order to estimate the total number of proton counts. Then, the contributions from weaker excited-state transitions identified in Fig. \ref{fig:backgroundSubtractionFits} were iteratively added to these fits. After correcting for SeGA's $\gamma$ efficiency, the positions and intensities of these smaller peaks were fixed in the cumulative spectrum. In energy regimes where $\beta$-delayed proton transitions to excited states are visible in the Fig. \ref{fig:backgroundSubtractionFits} $p_{\text{all}}$ spectrum, the number of decays determined by fitting the singles peaks agrees with the integrals from the fits of the $\gamma$-gated, background-subtracted spectra to within a factor of $\lesssim 2$ after accounting for SeGA's detection efficiency.   

In order to report normalized $\beta$-delayed proton intensities $I_p$ and plot $\Sigma B(\text{GT})$ as a function of $^{31}$S excitation energy, the Proton Detector efficiency must be quantified. Detailed simulations using the GEometry ANd Tracking (\textsc{Geant4}) framework were performed during the design, testing, and commissioning phases of GADGET development \cite{Friedman2019}. These Monte Carlo simulations were performed by randomly sampling the positions of decay events from an initial $^{31}$Cl beam distribution and counting the number of proton tracks that were contained within the active region of the Proton Detector. The longitudinal component of this beam distribution was modeled using the ionization drift times plotted in Fig. \ref{fig:longBeamDist}, while the transverse component was defined by the beam spot in Fig. \ref{fig:beamSpot}. Using the multiplicity of total counts detected in each of the five inner MICROMEGAS pads, a $\chi^2$-minimization procedure was used to determine the position and width of the Gaussian beam spot to within $\pm$6 mm. 

Since the completion of NSCL Experiment 17024, the Proton Detector has undergone an upgrade, allowing the GADGET II system to operate as a time projection chamber (TPC) at the Facility for Rare Isotope Beams (FRIB) \cite{Mahajan2024}. In the process of this upgrade, it was discovered that the degree to which electrons diffuse in P10 gas could have a larger effect on the final detection efficiency than originally assumed. Decay protons lose most of their kinetic energy in the Proton Detector through ionizing the gaseous medium, and the total number of electrons liberated during ionization depends on both the energy loss in the gas and the average energy required to create a single electron-ion pair for a given gas mixture. As the ionization electrons drift through the chamber under the influence of a uniform electric field, they diffuse transversely according to the relation $\sigma(t) = \sqrt{4D_et}$ until they reach the amplification region of the MICROMEGAS \cite{Sauli2014}. For an 800-Torr P10 gas mixture and a 150-V/cm drift field, the electron diffusion constant was calculated to be $D_e = 9.12(27) \times 10^{3}$ cm$^2$/s using the \textsc{MAGBOLTZ} program \cite{Biagi1999}.

To determine the size of this effect on the detection efficiency, we developed a Simplified Geometric Monte Carlo (\textsc{SGMC}) model that was able to reproduce the more detailed \textsc{Geant4} results, which do not account for electron diffusion, to within 3\% for proton energies up to 1.5 MeV \cite{Budner2026}. Using the same initial $^{31}$Cl beam spot and longitudinal profile, the \textsc{SGMC} model randomly samples the starting coordinates of a decay event from the reconstructed beam distribution, assumes proton radiation is emitted isotropically, and uses stopping powers calculated in \textsc{SRIM} to determine the final position of the thermalized proton within the gaseous detector medium \cite{Ziegler2010}. If a proton's full ionization track is contained within the active volume, we consider this a detected event; otherwise, this decay event triggers our veto condition. In order to incorporate the effect of electron diffusion into this model, for every 10-$\mu$m step along the length of the proton track, we randomly sample the positions of $n_{e^-}$ ionization electrons from a two-dimensional Gaussian distribution, whose size depends on the drift time of a randomly selected $\beta$-decay event, and where $n_{e^-}$ is proportional to the proton's $\Delta E$ over that 10-$\mu$m distance. Ultimately, we used the \textsc{SGMC} model to evaluate GADGET's efficiency across all relevant proton energies. Figure \ref{fig:pdEfficiency} compares simulation results with and without electron diffusion to the \textsc{Geant4} model.

\begin{figure}
\centering
\includegraphics[width=\linewidth]{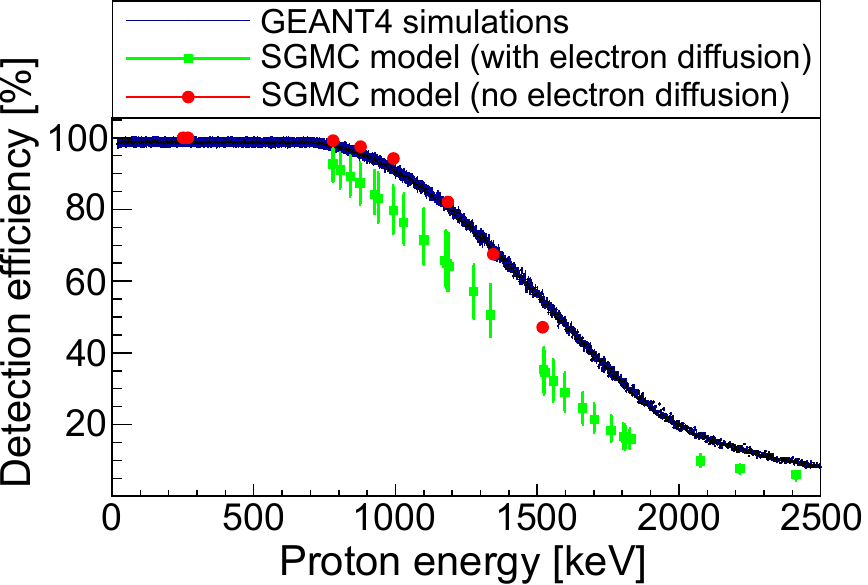}
\caption{\label{fig:pdEfficiency} Proton Detector efficiencies comparing \textsc{Geant4} simulations with no electron diffusion [solid black (blue online)] to the \textsc{SGMC} model, both with [square markers (green online)] and without [circular markers (red online)] the radial electron diffusion effect.}
\end{figure}

Many iterations of the \textsc{SGMC} calculations were performed over a variety of experimental conditions in order to rigorously investigate these systematic effects on the Proton Detector efficiency. One of the largest sources of uncertainty in our efficiency model was the initial beam spot distribution. This uncertainty arises from both the diffusion of thermalized beam particles via Brownian motion, as well as the $\chi^2$-minimization procedure used to determine the parameters of the Gaussian distribution, ultimately resulting in a $2-5\%$ uncertainty on the detection efficiency of a 1.0-MeV proton. Setting the simulation's charge deposition threshold in terms of $n_{e^-}$ for proton event detection was also a source of systematic uncertainty in our model. This is because all MICROMEGAS pads have slightly different gains, resulting in thresholds spanning $5-20$ keV, which changes the number of ionization electrons required to register an event or trigger the veto condition. This triggering threshold leads to a $3-4\%$ uncertainty in the 1.0-MeV proton detection efficiency. Smaller effects ($\lesssim$1\%) on the final error bars can be attributed to uncertainties in the \textsc{SRIM} model \cite{Ziegler2010}, the electron diffusion constant as calculated by \textsc{MAGBOLTZ} \cite{Biagi1999}, and the electric field distortions produced by slowly drifting positive ions generated by beam particles implanted at the end of the beam-delivery cycle, right before the start of the decay-measurement period.

The relative intensities $I_p$(rel) of all observed transitions along with their corresponding uncertainties are tabulated in Table \ref{table:allProtonData}. Just as in Ref. \cite{Budner2022}, the present $\beta$-delayed proton intensities were normalized relative to the 1.0-MeV decay intensity $I_p\text{(abs)}=1.31(2)\%$ adopted from literature \cite{Saastamoinen2011,Batchelder2020}. After refining GADGET's proton-detection efficiency model with more detailed numerical simulations, we report $I_p$(abs) = $9.0(10) \times 10^{-6}$ for the 258-keV resonance of interest. Thus, the corresponding proton branching ratio and resonance strength are $\Gamma_p/\Gamma = 2.4(3) \times 10^{-4}$ and $\omega\gamma=77(45)$ $\mu$eV, respectively. These revised numbers supersede the previously reported values, are well within the reported uncertainties, and any discrepancies have a negligible effect on the astrophysical conclusions.

Table \ref{table:allProtonData} shows that several $^{31}$S excited states populated in $^{31}$Cl decay are located above the $\alpha$ emission threshold at $S_{\alpha}=9082.94(25)$ keV \cite{Wang2021}, but the calorimetric nature of GADGET prevents distinguishing $\alpha$ particles from protons. Discriminating between $^{31}$Cl $\beta$-delayed proton and $\alpha$ events is now possible with the recent TPC upgrade to the GADGET II system \cite{Mahajan2024}. However, since the Coulomb barrier suppresses significant $\alpha$ emission for charged-particle decays $<10$ MeV, we assume all $\beta$-delayed particles observed in our spectra are the result of proton decays.

\begin{table*}[htp]
\renewcommand{\arraystretch}{1.5}
{\setlength{\tabcolsep}{0.75pt}
\begin{tabular}{cc|c|c|c|c|c|cc|cc} 

\hline
\multicolumn{2}{c|}{$E_{x_i}$($^{31}$S) [keV]} & $J_i^{\pi}$($^{31}$S) & \multicolumn{1}{c|}{$I_{\beta}$(abs) [\%]} & \multicolumn{1}{c|}{log($ft$)} & \multicolumn{1}{c|}{$E_{x_f}$($^{30}$P) [keV]} & $J_f^{\pi}$($^{30}$P) & \multicolumn{2}{c|}{$E_p$(c.m.) [keV]} & \multicolumn{2}{c}{$I_p$(rel) [\%]} \\

\cite{NuDat31S} & present & present & present & present & \cite{NuDat30P} & \cite{NuDat30P} & \cite{Saastamoinen2011} & present & \cite{Saastamoinen2011} & present \\
\hline
\hline
6390.46(16) & 6388(4) & $3/2^+$ \cite{Bennett2016} & 3.38(18) \cite{Bennett2016} & $4.11(2)$ & 0 & $1^+$ & --- & 258(4) & --- & 0.069(8)  \\
\hline
6936(2) & --- & $\bm{1/2^+},\bm{3/2^+},5/2^+$ & 0.20(3) & $5.09^{+0.07}_{-0.06}$ & 0 & $1^+$& 806(2) & $^\dagger$805(2) & 20.4(2) & $15.4(15)$ \\
\hline
7037(2) & --- & $5/2^+$ \cite{Burcher2022} & 0.087(14) & $5.40(7)$ & 0 & $1^+$ & 906(2) & $^\dagger$906(2) & 12.4(2) & 6.6(7) \\
\hline
7157.5(11) & --- & $\bm{3/2^+},5/2^+$ \cite{Setoodehnia2020,Burcher2022} & 1.31(2) \cite{Saastamoinen2011} & $4.162(7)$ & 0 & $1^+$ & 1026(2) & $^\dagger$1026.9(11) & 100(4) & $100(12)$ \\
\hline
\multirow{2}{*}{7699(3)} & \multirow{2}{*}{7703(3)} & \multirow{2}{*}{$\bm{1/2^+},\bm{3/2^+},5/2^+$} & \multirow{2}{*}{0.14(3)} & \multirow{2}{*}{$4.83^{+0.12}_{-0.09}$} & 708.70(7) & $1^+$ & --- & 870(18) & --- & $<0.04$ \\
        &      &   & & & 0 & $1^+$  & 1571(3) & 1572(3) & 21.0(4) & 10(2) \\
\hline
7774(3) & *7797(12) &  & $<0.0005$ & $>7.18$ & 708.70(7) & $1^+$ & --- & 958(12) & --- & $<0.05$ \\
\hline
7895(2) & 7887(3) & $\bm{1/2^+},\bm{3/2^+},5/2^+$ & $0.038^{+0.010}_{-0.009}$ & $5.29^{+0.11}_{-0.10}$ & 0 & $1^+$ & 1763(3) & 1756(3) & 6.4(2) & $2.9^{+0.7}_{-0.6}$ \\
\hline
7907(3) & 7905(8) & $1/2^+$ \cite{Ma2007} & 0.0027(8) & $6.43^{+0.16}_{-0.12}$ & 708.70(7) & $1^+$ & --- & 1065(8) & --- & 0.20(6) \\
\hline
8015(3) & 8017(3) & $\bm{1/2^+},\bm{3/2^+},5/2^+$ & 0.09(2) & $4.83^{+0.11}_{-0.10}$ & 0  & $1^+$      & 1891(3) & 1887(3) & 10.9(2) & $6.9^{+1.5}_{-1.3}$ \\
\hline
8022(3) & 8024(2) & $1/2,3/2,5/2$ & $0.0038^{+0.0013}_{-0.0012}$ & $6.20^{+0.17}_{-0.13}$ & 677.11(10) & $0^+$ & --- & 1217(2) & --- & 0.29(9) \\
\hline
8268(10) & 8275(4) &  & $<0.007$ & $>5.76$ & 0  & $1^+$   & 2139(17) & 2145(4) & 1.3(1) & $<0.6$ \\
\hline
\multirow{4}{*}{8424(3)} & \multirow{4}{*}{8418(3)} & \multirow{4}{*}{$\bm{1/2^+},3/2^+,5/2^+$} & \multirow{4}{*}{0.09(2)} & \multirow{4}{*}{$4.53^{+0.11}_{-0.09}$} & 1454.28(5) & $2^+$ & --- & $^\dagger$833(3) & --- & 0.105(19) \\
        &     &    & & & 708.70(7) & $1^+$ & --- & $^\dagger$1579(3) & --- & 1.0(2) \\
        &     &   & & & 677.11(10) & $0^+$ & --- & $^\dagger$1610(3) & --- & $2.2^{+0.6}_{-0.5}$ \\
        &     &    & & & 0         & $1^+$ & 2298(3) & $^\dagger$2287(5) & 2.3(1) & $4.0^{+1.5}_{-1.4}$ \\
\hline
\multirow{2}{*}{8563(2)} & \multirow{2}{*}{8557(2)} & \multirow{2}{*}{$1/2^+,\bm{3/2^+},5/2^+$} & \multirow{2}{*}{0.015(3)} & \multirow{2}{*}{$5.21^{+0.10}_{-0.08}$} & 1454.28(5) & $2^+$ & --- & 973(2) & --- & 0.33(6) \\
        &    &     & & & 708.70(7) & $1^+$ & --- & 1715(4) & --- & 0.8(2) \\
\hline
\multirow{2}{*}{8702(17)} & *8669(8) &  & $<0.0019$ & $>6.04$ & 677.11(10) & $0^+$ & --- & 1862(8) & --- & $<0.14$ \\
                          & 8708(7) & $(\bm{1/2^+},\bm{3/2^+},5/2^+)$ & $0.0028^{+0.0010}_{-0.0009}$ & $5.82^{+0.16}_{-0.13}$ & 708.70(7) & $1^+$ & --- & 1869(7) & --- & $0.22^{+0.07}_{-0.06}$ \\
\hline
8815(3) & 8813(2) & $1/2^+,\bm{3/2^+},\bm{5/2^+}$ & 0.020(6) & $4.88^{+0.15}_{-0.12}$ & 1454.28(5) & $2^+$ & 1225(3) & 1228(2) & 2.7(1) & $1.6^{+0.5}_{-0.4}$ \\
\hline
\multirow{2}{*}{8973(4)} & *8950(5) & $\bm{1/2^+},3/2^+,5/2^+$ & 0.010(8) & $5.1^{+0.8}_{-0.3}$ & 677.11(10) & $0^+$ & --- & 2143(5) & --- & $0.7(6)$ \\ 
                         & 8964(2) & $1/2^+,\bm{3/2^+},\bm{5/2^+}$ & 0.009(2) & $5.10^{+0.11}_{-0.10}$ & 1454.28(5) & $2^+$ & 1390(17) & 1379(2) & 1.3(12) & $0.68^{+0.15}_{-0.13}$ \\
\hline
\multirow{2}{*}{9226(25)} & \multirow{2}{*}{9238(3)} & \multirow{2}{*}{$1/2^+,3/2^+,\bm{5/2^+}$} & \multirow{2}{*}{0.016(4)} & \multirow{2}{*}{$4.58^{+0.12}_{-0.09}$} & 1973.34(8) & $3^+$ & --- & 1137(4) & --- & $0.38(17)$ \\
        &    &     & & & 1454.28(5) & $2^+$ & 1647(17) & 1651(3) & 1.4(2) & 0.8(2) \\
\hline
\multirow{2}{*}{9332(30)} & \multirow{2}{*}{9328(5)}& \multirow{2}{*}{$1/2^+,3/2^+,5/2^+$} & \multirow{2}{*}{$0.017^{+0.007}_{-0.008}$} & \multirow{2}{*}{$4.44^{+0.25}_{-0.15}$} & 1973.34(8) & $3^+$ & --- & 1223(5) & --- & $0.32^{+0.09}_{-0.08}$ \\
         &     &    & & & 708.70(7) & $1^+$ & --- & 2489(8) & --- & $1.0^{+0.5}_{-0.6}$ \\
\hline
\multirow{3}{*}{9423(7)} & \multirow{2}{*}{*9405(4)} & \multirow{2}{*}{$1/2^+,3/2^+,\bm{5/2^+}$} & \multirow{2}{*}{$0.0033^{+0.0011}_{-0.0009}$} & \multirow{2}{*}{$5.08^{+0.14}_{-0.12}$} & 2538.87(18) & $3^+$ & --- & 740(10) & --- & 0.0027(9) \\
                         &  &  &  &  & 1454.28(5) & $2^+$ & --- & 1819(6) & --- & $0.25^{+0.08}_{-0.06}$ \\
                         & 9421(8) & $3/2^+,5/2^+$ \cite{Ma2007} & 0.004(2) & $4.99^{+0.30}_{-0.18}$ & 1973.34(8) & $3^+$ & --- & 1317(8) & --- & 0.29(14) \\
\hline
9777(2) & *9730(10) & & $<0.0005$ & $>5.46$ & 2538.87(18) & $3^+$ & --- & 1060(10) & --- & $<0.04$ \\

\end{tabular}}
\caption{Comprehensive results from the $^{31}$Cl($\beta p \gamma$)$^{30}$P measurement. Present initial excitation energies $E_{x_i}$ and center-of-mass decay energies $E_p$ are compared to the nearest nuclear levels evaluated in literature and previously measured $\beta$-delayed proton energies, respectively.  Level energies denoted by an asterisk (*) have not been previously reported, and proton-decay energies denoted by a dagger ($\dagger$) were used for energy calibration. When available, established spin-parity assignments are adopted from literature. Otherwise, $J_i^{\pi}$ values are deduced from log($ft$) calculations, and those shown in bold assume a purely $\ell=0$ transition. Levels where only log($ft$) lower limits are reported receive no assignment. The log($ft$) values for $^{31}$S states were computed using the \textsc{RadiationReport} software as part of the Evaluated Nuclear Structure Data File Analysis and Utility Programs \cite{Chen2025}. $I_{\beta}$ represents the number of times a $^{31}$S level is directly populated for every 100 $\beta^+$ decays or electron captures; most of the $\beta^+$ feeding to the 6390-keV excited state is observed through $\gamma$ decay \cite{Bennett2016,Bennett2018}. $I_p$ values are normalized relative to the intensity of the 1.0-MeV proton decay reported in Ref. \cite{Saastamoinen2011,Batchelder2020}. The final excitation energies $E_{x_f}$ for levels directly populated by $^{31}$Cl $\beta$-delayed proton decay are also tabulated. All derived quantities were computed using full-precision values; reported values were rounded only at the final step to match the precision implied by the quoted uncertainties.}
\label{table:allProtonData}
\end{table*}

\subsection{D. Beam contaminant effects on uncertainties}

Although no $\beta$-delayed particle emitters were initially identified as primary beam contaminants from the $\Delta E$-TOF spectrum shown in Fig. \ref{fig:particleID}, the identification of a 1779-keV photopeak in Fig. \ref{fig:protonGatedGammas} suggests that at least a small amount of $^{29}$S contamination was present in our radioactive beam data. This $\gamma$ ray is emitted from the decay of the first excited state of $^{28}$Si, which is populated by $^{29}$S $\beta$-delayed proton decay \cite{Viera1979,Zha2026}. Gating on these $\gamma$ events reveals the presence of 766- and 1302-keV protons in our dataset. There are $\approx 10$ proton transitions for which $0.7<E_p<3.0$ MeV that could be present in our spectrum as a result of  $^{29}$S decay. However, due to the very limited statistics of observable events, we used the total number of 766- and 1302-keV protons detected, our evaluated Proton Detector and SeGA efficiency models, as well as the relative proton-decay intensities tabulated in Ref. \cite{Batchelder2020} to estimate the magnitude of $^{29}$S decays contributing to our experimental dataset.

The effects of $^{29}$S contamination on the relative uncertainties in $I_p$(rel) for $^{31}$Cl $\beta$-delayed proton decays observed within the energy regions  $0.7<E_p<1.4$ MeV, $1.5<E_p<2.0$ MeV, and $2.0<E_p<2.7$ MeV are 0.18(4)\%, 0.07(1)\%, and 4.8(17)\%, respectively. Applying this same logic to the expected contributions of previously reported $^{31}$Cl $\beta$-delayed proton decays, specifically from the 2.0-, 2.4-, and 2.6-MeV proton transitions which were not observed in our spectra \cite{Saastamoinen2011,Batchelder2020}, increases our relative uncertainties in decay intensity to 5.7(8)\% and 32(6)\% for  proton energies $1.5<E_p<2.0$ MeV and $2.0<E_p<2.7$ MeV, respectively. All other possible $\beta$-delayed proton emitters within the $A\approx30$ mass region were also investigated as possible beam contaminants, but none were definitively identified in our data, and their contributions to our proton spectra should be considered negligible.
\section{V. Results and Discussion}

\begin{figure}
\centering
\includegraphics[width=\linewidth]{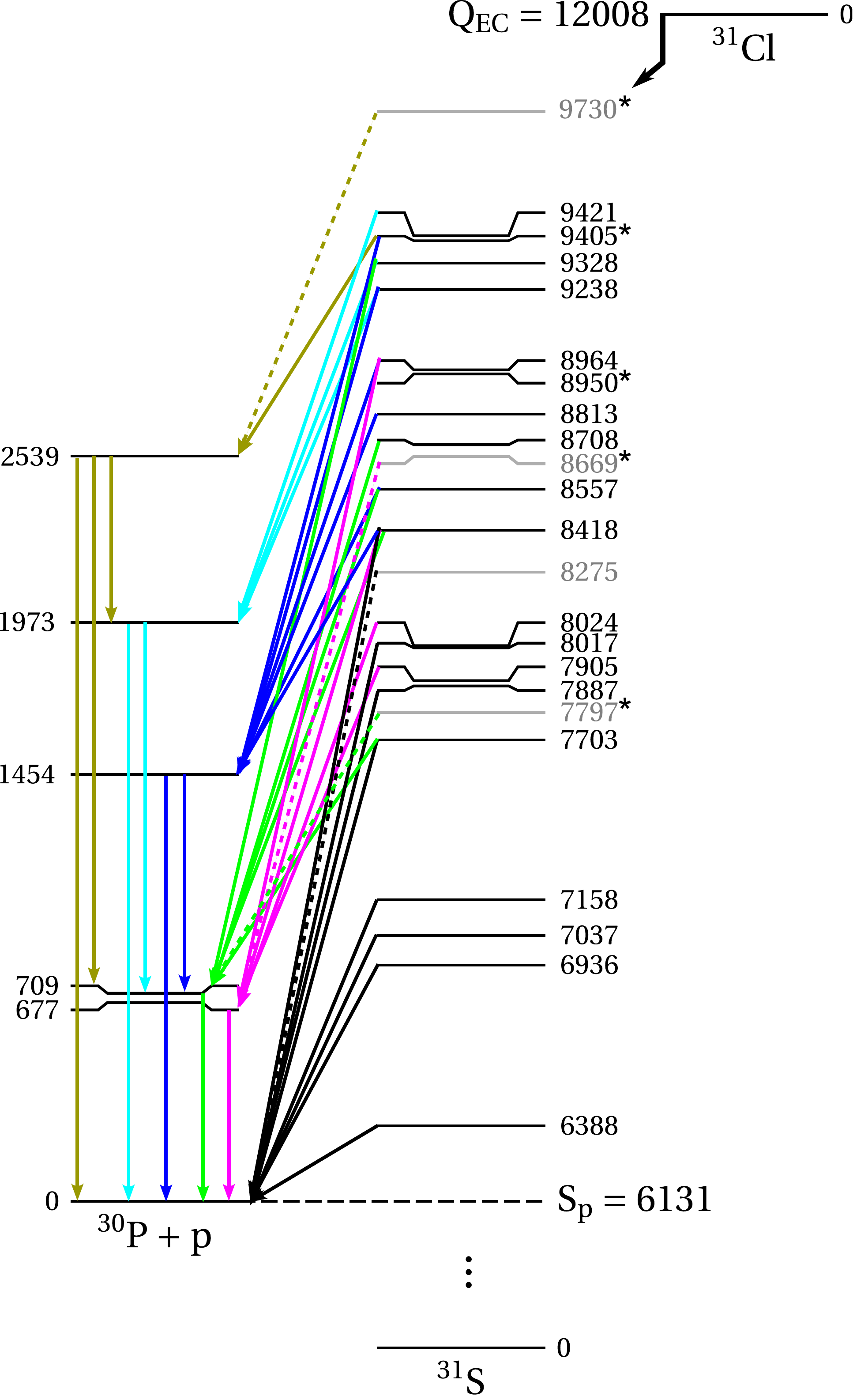}
\caption{The $^{31}$Cl($\beta p \gamma$)$^{30}$P decay scheme. Transitions from proton-unbound levels in $^{31}$S to excited states in $^{30}$P are shown as gray arrows [magenta ($p_1$), green ($p_2$), blue ($p_3$), cyan ($p_4$), and gold ($p_5$) online], while ground-state ($p_0$) transitions are represented by black arrows. The $^{30}$P $\gamma$-ray transitions are color-coded to match the $\beta$-delayed proton decays populating their respective excited states. The arrows of tentative transitions are dashed, while tentative $^{31}$S level assignments are shown in gray. Previously unreported levels are denoted by an asterisk (*). All energies are reported in units of keV.}
\label{fig:BetaDecayScheme}
\end{figure}

In total, as many as 31 unique $\beta$-delayed proton transitions have been observed in NSCL Experiment 17024, of which 11 have been reported in prior $^{31}$Cl $\beta$-decay studies \cite{Aysto1982,Aysto1983,Aysto1985,Ognibene1996,Kankainen2006,Saastamoinen2011,Kankainen2014,Saastamoinen2016}. Although all previously observed decays were reported to populate the ground state of $^{30}$P, up to 20 of the present transitions are now assigned to one of the five lowest-lying excited states of $^{30}$P, as illustrated in Fig. \ref{fig:BetaDecayScheme}. In our discussion of these experimental findings, we argue for the existence of two new $^{31}$S proton-unbound levels, provide constraints on spin-parity assignments, and compare our extracted $B(\text{GT})$ distribution to both theory and past experimental results.

\subsection{A. Excitation energies for $^{31}$S levels}

Almost all of the  $^{31}$S excitation energies determined in this work agree with the most recently evaluated nuclear level assignments to within two standard deviations \cite{NuDat31S}. Therefore, it is reasonable to conclude that the vast majority of states populated in $^{31}$Cl $\beta^+$ decay have already been observed in previous experiments using various techniques and channels. Nevertheless, there are several instances where the present excitation energies disagree with prior measurements by more than two standard deviations.

In most cases, any evidence for potentially new $^{31}$S excited states is based solely on the observation of small excesses within individual $\gamma$-gated proton spectra. This applies to the tentative level assignments of 7797, 8669, and 9730 keV, which are depicted in gray within the level-scheme diagram of Fig. \ref{fig:BetaDecayScheme}. These excitation energies were deduced using Eq. \ref{eq:exEn} from the tentative observations of 958-, 1862-, and 1060-keV proton decays in the $p_2$, $p_1$, and $p_5$ spectra of Fig. \ref{fig:backgroundSubtractionFits}, respectively. However, the large uncertainties associated with performing background subtraction in the 1.0-MeV energy region prevent conclusive identification of new proton decays and $^{31}$S excited states from the $p_2$ and $p_5$ spectra alone. In the case of the $p_1$ spectrum, the excess counts around $E_p\approx1.9$ MeV could be the result of incomplete background subtraction due to the presence of real coincidences between $p_2$ transitions and the low-energy tail of the nearby 709-keV photopeak. Therefore, we consider the observation of all three proton decays and their corresponding $^{31}$S excited states to be tentative.

The other $^{31}$S level for which there is potential, albeit inconclusive, evidence for $\beta^+$ feeding is the 8275-keV excited state. This level has been previously measured in a $^{31}$P($^3$He,$t$)$^{31}$S reaction experiment \cite{Wrede2007,Wrede2009} and was assumed to be the origin of the 2.1-MeV $\beta$-delayed $p_0$ decay reported in Ref. \cite{Saastamoinen2011}. In the present work, we have identified the existence of a $p_1$ transition at $E_{p_1}=2143(5)$ keV, which would imply a 677-keV-higher excitation energy. However, the number of 2.1-MeV counts observed in the proton singles spectrum is $\approx1.7$ times greater than the number of $\gamma$-efficiency-corrected counts determined from the $p_1$ spectrum fit. Thus, it is still possible that both a 2145-keV $p_0$ transition and a 2143-keV $p_1$ transition are present in our data. For this reason, we consider $\beta^+$ feeding to the 8275-keV state via $^{31}$Cl decay to be tentative but argue that the 2143-keV transition to the first excited state of $^{30}$P constitutes evidence in favor of a new $^{31}$S level, for which  $E_{x_i}=8950(5)$ keV.

Furthermore, we report two additional $\beta$-delayed proton transitions to $^{30}$P excited states which suggest the existence of another previously unobserved $^{31}$S level. First, a weak 1819-keV transition to the 1454-keV excited state of $^{30}$P can be seen in the $p_3$ spectrum of Fig. \ref{fig:backgroundSubtractionFits}. Second, the $0.7<E_p<0.8$ MeV region of Fig. \ref{fig:hLevel5protons} contains excess charged-particle events detected in coincidence with 2539-keV $\gamma$ rays. Using Eq. \ref{eq:exEn} and taking a weighted average, we calculate the excitation energy of this new $^{31}$S level to be $E_{x_i}=9405(4)$ keV, whose decays populate the third and fifth excited states of $^{30}$P.

In summary, we present evidence for the existence of two new $^{31}$S levels at 8950 and 9405 keV and tentatively report the observation of proton decays from previously unobserved $^{31}$S excited states at 7797, 8669, and 9730 keV. We acknowledge the possibility of $\beta^+$ feeding to a known level at $E_{x_i} \approx 8275$ keV but also consider this population to be tentative. All other excitation energies reported at present should be interpreted as independent measurements of previously observed levels.

\begin{figure}
\includegraphics[width=\linewidth]{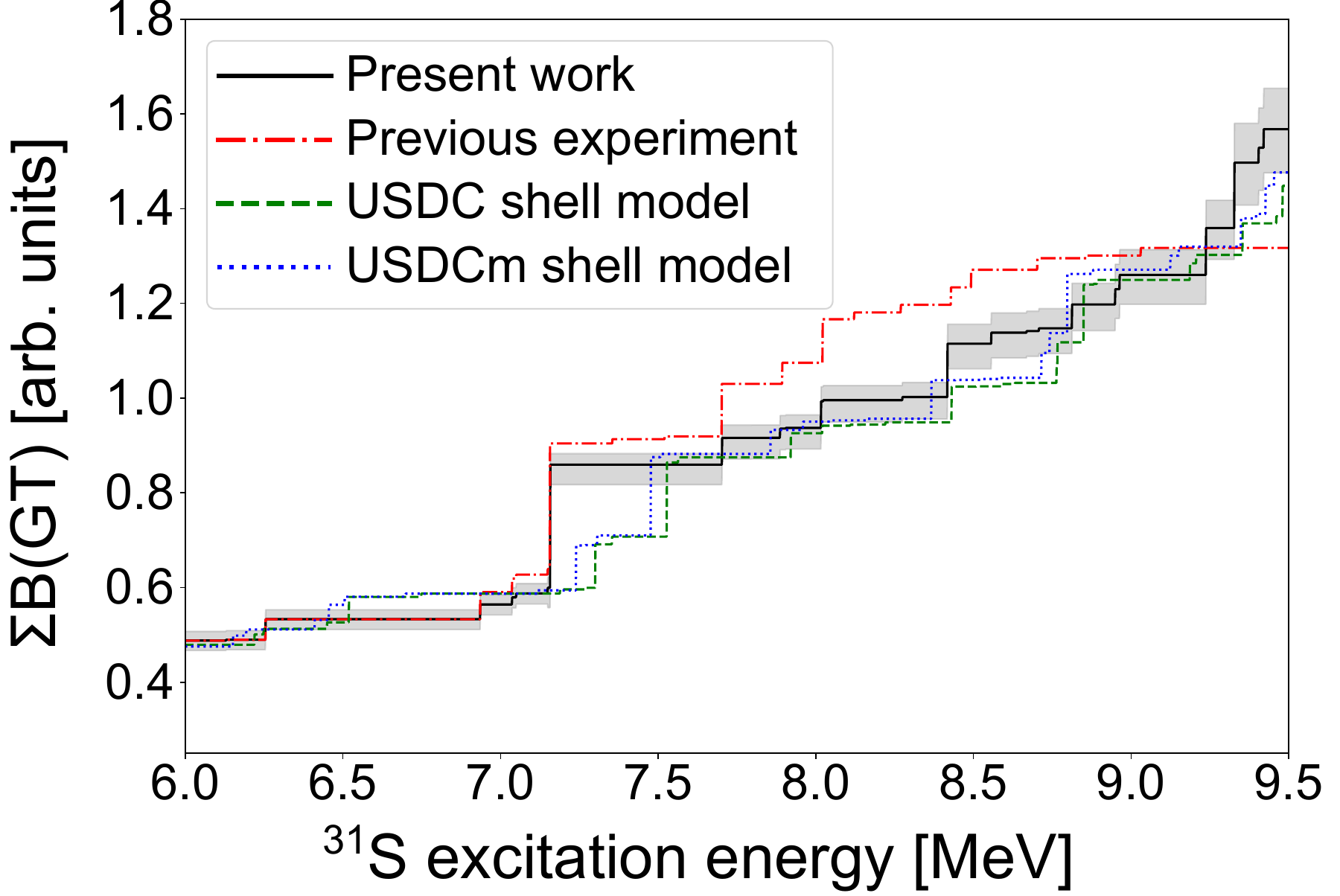}
\caption{The cumulative GT strength distribution for $^{31}$S proton-unbound states $6.0<E_{x_i}<9.5$ MeV. The present experimental evaluation [solid black] is compared to prior measurements [dash-dotted (red online)] from Refs. \cite{Saastamoinen2011}, as well as to shell-model theory calculations using the USDC [dashed (green online)] and USDCm [dotted (blue online)] Hamiltonians \cite{Magilligan2020}. The gray uncertainty band was determined using propagation of errors for the log($ft$) values in Table \ref{table:allProtonData} as well as the constants $K/g^2_V$ and $g_A/g_V$ from Refs. \cite{Hardy2020,Navas2024}.}
\label{fig:SumBGT}
\end{figure}

\subsection{B. Spin-parity arguments}

The spin-parity of the $^{31}$Cl ground state is $J^{\pi}=3/2^+$, and the allowed angular momentum changes according to $\beta$-decay selection rules are $\Delta J = 0, \pm 1$, with no change in parity ($\Delta \pi=0$). Thus, in general, the $^{31}$S levels populated in $^{31}$Cl $\beta^+$ decay are constrained to be $J_i^{\pi} = 1/2^+, 3/2^+, 5/2^+$. When available, established spin-parity assignments are adopted from existing literature on the basis of symmetry arguments for the $^{31}$P mirror nucleus \cite{Bennett2016} and angular distribution data from single-nucleon transfer reaction measurements \cite{Ma2007,Setoodehnia2020,Burcher2022}. Otherwise, the $J_i^{\pi}$ assignments reflected in Table \ref{table:allProtonData} are deduced from the log($ft$) calculations in accordance with the National Nuclear Data Center policy. These guidelines for $\beta$ decay are outlined in Refs. \cite{Raman1973,Turkat2023,NDS2025}.

Looking at Fig. \ref{fig:BetaDecayScheme}, a keen observer might notice that ground-state proton transitions only appear to be emitted from $^{31}$S levels where $E_{x_i}<8.5$ MeV. For example, the 8557-keV level decays to the 709-keV, $J_f^{\pi}=1^+$ and 1454-keV, $J_f^{\pi}=2^+$ excited states of $^{30}$P. Therefore, it seems likely that the ground state of $^{30}$P would also be populated by the decay of this proton-unbound resonance, since it is more energetically favorable and is also a $J_f^{\pi}=1^+$ level. However, due to the geometry of the Proton Detector's active region and the chosen gas pressure for this experiment, the efficiency to detect protons with $E_p>2.0$ MeV is $\lesssim10\%$. Combined with potential contributions from multiple proton transitions, limited energy resolution, and without the $\gamma$-tagging capabilities on excited state decays to apply coincidence gates, definitively distinguishing between higher-energy $p_0$ transitions and continuum background becomes increasingly challenging as a function of decay energy within the present experimental configuration.

\subsection{C. Comparing theory to experiment}

Comparing the present work with data from the best prior measurement of $^{31}$Cl $\beta$-delayed proton decay, we have identified 11 of the same peaks reported by Saastamoinen, and our relative proton-decay intensities agree within a factor of $\lesssim2$. In all but one case, the present intensities are lower than those previously reported. In Ref. \cite{Saastamoinen2011}, the summed total of all $^{31}$Cl $\beta$-delayed proton decays yields an absolute intensity of $I_{\beta p}=2.4(2)\%$. Our present result of $I_{\beta p}=2.08(6)\%$ agrees to within 1.5 standard deviations of the previous experimental findings. The discrepancies between this work and that of Ref. \cite{Saastamoinen2011} are visualized in Fig. \ref{fig:SumBGT} when plotting cumulative $B(\text{GT})$ values. The Gamow-Teller strength has been evaluated for each $^{31}$S level using the expression 

\begin{equation}
B(\text{GT}) = \frac{K/g^2_V}{ft(g_A/g_V)^2},
\label{eq:BGT}
\end{equation}

\noindent
where our computed $ft$ values from Table \ref{table:allProtonData} are combined with the constants $K/g^2_V = 6144.48(370)$ s and $g_A/g_V = -1.2753(13)$ from Refs. \cite{Hardy2020,Navas2024}, respectively. Our calculations ignore the negligible Gamow-Teller contribution to the predominantly $T=3/2$, 6279-keV isobaric analog state, for which proton decay is isospin forbidden. They also neglect the mostly $T=1/2$, proton-emitting level at 6388 keV, whose $\beta^+$ feeding is enhanced by the Fermi interaction due to isospin mixing with the isobaric analog state \cite{Bennett2016}. We compare our experimental evaluation to the universal $sd$-shell model using the isospin-breaking Hamiltonians: USDC and USDCm \cite{Magilligan2020}. The theoretical results plotted in Fig. \ref{fig:SumBGT} adopt a quenching factor of $q_{\text{GT}}=0.764$ from Table III in Ref. \cite{Richter2008}. Our findings are in better agreement with shell-model theory than previous work for excitation energies $7.5<E_{x_i}<9.5$ MeV. This results from the $\beta$-feeding intensities spreading out over a larger range of excitation energies, as opposed to being concentrated within a smaller number of proton-unbound levels, located in relatively close proximity to the proton emission threshold of $^{31}$S, which were originally assumed to proton-decay directly to the ground state of $^{30}$P.

\subsection{D. Future considerations}

The present experimental setup was designed to measure small proton branching ratios of low-energy resonances that dominate the $^{30}$P($p,\gamma$)$^{31}$S reaction rate during the thermonuclear runaway event in oxygen-neon novae. This required high-efficiency detection of low-energy, $\beta$-delayed proton decays with minimal $\beta^+$ background. In order to extract more information on $^{31}$S levels for which $E_{x_i}>9.5$ MeV, the pressure of the Proton Detector's fill gas could be increased up to $\approx2000$ Torr, which would improve detection efficiency for higher decay energies at the expense of larger $\beta^+$ backgrounds and degraded energy resolution \cite{Dopfer2026}. The TPC capabilities of GADGET II could be utilized to distinguish between $\beta$-delayed proton and $\alpha$ decays \cite{Mahajan2024}. Alternatively, a $\Delta E$-$E$ silicon telescope could be deployed in tandem with a high-efficiency HPGe array to measure higher-energy, $\beta$-delayed charged-particle decays in coincidence with $^{30}$P excited-state $\gamma$ decays. Although low-energy $\alpha$ particles would struggle to penetrate through the thin $\Delta E$ detector, they could be identified by vetoing events that reach the thicker $E$ detector. Employing this technique would improve both energy resolution and detection efficiency for charged-particle energies $E_p>1.0$ MeV.
\section{VI. Conclusions}

This study provides the first detailed investigation of the $^{31}$Cl($\beta p \gamma$)$^{30}$P decay sequence. Previous measurements of $^{31}$Cl $\beta^+$ decay only provided substantial information on the proton singles spectrum or $\gamma$-ray emission from $^{31}$S excited states. The high-statistics proton-$\gamma$ coincidence data collected at NSCL during this radioactive beam experiment reveal many previously unobserved $\beta$-delayed proton transitions to excited states in the $^{30}$P daughter nucleus. Our analysis led to the identification of at least two new $^{31}$S proton-unbound levels. The present findings demonstrate an example of $\beta$-delayed ``proton pandemonium,'' and the techniques discussed throughout this article are broadly applicable to constructing accurate $\beta$-decay schemes for comparison with theoretical models.

Furthermore, this work demonstrates the sensitivity of the GADGET system for measuring weak, low-energy, $\beta$-delayed charged-particle decays in coincidence with high-efficiency, $\gamma$-ray detection capabilities and good energy resolution. GADGET has already been successfully utilized in several published scientific experiments \cite{Friedman2020,Sun2021,Budner2022} and is a useful instrument for studying resonances involved in charged-particle capture on both ground and excited states of unstable nuclei in astrophysical environments. Our experimental results demonstrate that it can serve as a sensitive probe of nuclear structure as well. The recent upgrade to GADGET II integrates the gaseous detector system with the FRIB Decay Station Initiator \cite{Mahajan2024}, enabling even more detailed and sensitive measurements in the future with both the Proton Detector (GADGET) and TPC (GADGET II) configurations.

Research at NSCL was funded by the National Science Foundation under Grant Nos. PHY-1913554, PHY-1102511, PHY-1565546, PHY-1811855, PHY-2310059, PHY-2110365, PHY-2209429, and PHY-2514797 as well as by the Department of Energy Office of Science under Award Nos. DE-SC0016052 and DE-SC0024587. We acknowledge support from the Natural Sciences and Engineering Research Council of Canada, the Spanish MINECO Grant AYA2017-86274-P, FEDER resources, the AGAUR/Generalitat de Catalunya Grant No. SGR-661/2017, and EU Horizon 2020 Grant No. 101008324 (ChETEC-INFRA). This article also benefited from discussions within the ChETEC COST Action (CA16117). Additional funding sources include National Research Foundation of Korea Grant No. RS-2024-00338255. We thank the NSCL staff for providing technical and administrative support for this experiment.

\endgroup

\bibliography{31Cl_decay}

\end{document}